\documentclass[aps,pre,twocolumn]{revtex4-1}
\usepackage{amssymb}
\usepackage{amsmath,bm}
\usepackage{graphicx,color}
\usepackage{epsfig}
\usepackage{undertilde}
\usepackage{multirow}
\usepackage[cal=pxtx]{mathalfa}
\usepackage{accents}
\DeclareMathAccent{\wtilde}{\mathord}{largesymbols}{"65}

\usepackage{stackengine}
\stackMath
\newcommand\tenq[2][1]{%
	\def\useanchorwidth{T}%
	\ifnum#1>1%
	\stackunder[0pt]{\tenq[\numexpr#1-1\relax]{#2}}{\scriptscriptstyle\sim}%
	\else%
	\stackunder[1pt]{#2}{\scriptscriptstyle\sim}%
	\fi%
}
\renewcommand{\tensor}[1]{{\underline{\underline{#1}}}}
\renewcommand{\vec}[1]{{\underline{#1}}}
\newcommand{\B}[1]{{\underline{#1}}} 
\def\d {{\rm d}}
\def\uvec#1{{\hat{\underline{#1}}}}
\def\utilde#1{\underaccent{\wtilde}{#1}}
\def\uttilde#1{\underaccent{\wtilde}{\underaccent{\wtilde}{#1}}}
\begin{document}

\title{Stress Correlations in Frictional Granular Media}

\author{Ana\"el Lema\^{\i}tre$^1$, Chandana Mondal$^2$, Itamar Procaccia$^{2,3}$ and Saikat Roy$^2$}
\affiliation{$^1$NAVIER, UMR 8205, \'Ecole des Ponts ParisTech, IFSTTAR, CNRS, UPE, Champs-sur-Marne, France\\$^2$Department of Chemical Physics, the Weizmann Institute of Science, Rehovot 76100, Israel. \\$^3$Center for OPTical IMagery Analysis and Learning, Northwestern Polytechnical University, Xi'an, 710072 China}

\begin{abstract}
This paper investigates whether in frictional granular packings, like in Hamiltonian amorphous elastic solids, the stress
autocorrelation matrix presents long range anisotropic contributions just as elastic
Green's functions. We find that in a standard model of frictional granular packing this is not the case.  We prove quite generally
that mechanical balance and material isotropy constrain the stress auto-correlation matrix to be fully determined by two spatially isotropic functions: the pressure
and torque auto-correlations. The pressure and torque fluctuations being respectively normal and hyper-uniform
force the stress autocorrelation to decay as the elastic Green's function. Since we find the torque fluctuations to be hyper-uniform, the culprit is
the pressure whose fluctuations decay slower than normally as a function of the system's size. Investigating the reason for these abnormal pressure fluctuations we discover that anomalous correlations build up already during the compression of the dilute system before jamming. Once jammed
these correlations remain frozen. Whether this is true for frictional matter in general or is it the
consequence of the model properties is a question that must await experimental scrutiny and possible alternative models.
\end{abstract}

\maketitle

\section{Introduction}

During the last decade it became clear that the stress field of amorphous solids whose inter-particle forces derive from a Hamiltonian present long ranged correlation
tails of a form similar to elastic Green's functions \cite{09HC,14Lem,15Lem,17Lem,18Lem}. The first observations of this phenomenon in non-frictional granular media were viewed as evidence to Edward's ansatz \cite{89EO} about
the distribution of possible packing near the jamming point. But more recently it was demonstrated that these
long range correlations follow in Hamiltonian problems from the conjunction of three properties. These are
(i) Mechanical balance, (ii) Material isotropy and (iii) the normality of local pressure fluctuations \cite{17Lem,18Lem}. The derivation of these results depends crucially on the symmetry of local stress which
inevitably breaks down in the presence of frictional forces which introduce local torques. The question is then
fully open about the nature of stress correlations in frictional granular packings, an important, diverse and widespread class of materials including sand, soils, powders etc.

In Hamiltonian systems with central forces, mechanical balance and material isotropy demand the stress auto-correlation matrix to be fully determined by the pressure auto-correlation only. Here we show that in  frictional granular packings, in sharp contrast, it is determined not by one but by two spatially isotropic
functions, the pressure and torque autocorrelations. We will demonstrate that in the absence of external
torques, the torque fluctuations are hyper-uniform, i.e. the torque auto-correlation vanishes in the
zero wave-number limit. As a consequence the torque contribution to the stress auto-correlation is
sub-dominant at large wave-length. Consequently, the large distance decay of the stress-autocorrelation
is again determined by the scaling of local pressure fluctuations on domains of increasing sizes. When these
fluctuations are normal the presence of elastic-like long-ranged anisotropic contributions follows. We find however that
the pressure fluctuations are {\em not} normal, and the tails of the stress auto-correlation differ from those expected
in elastic systems, falling off more slowly.

The theoretical discussion in this paper will be backed by numerical simulations using the standard and time-honored
Cundall-Strack model \cite{79CS} of assemblies of frictional disks. The model is described briefly in Sect.~\ref{model}.
While this model has been used by hundreds if not thousands of researchers, it is a coarse grained model and our
conclusions regarding the nature of stress auto-correlation functions are achieved subject to the assumptions
embedded in it. Thus the final conclusion regarding how stress auto-correlation function decay in frictional granular matter at large distances
must await either experiments or other simulations using different models.

The next section \ref{theory} develops the theory of stress correlations in frictional
assemblies of disks. These purely theoretical results are expected to be independent of the particular coarse grained
model employed to simulate frictional granular matter. The main conclusion of the theory is that the decay of stress
correlations at large distance are determined by mechanical balance, material isotropy and the nature of torque and pressure fluctuations.
In Sect.~\ref{numerics} we present numerical simulations of the Cundall-Strack model and a demonstration of the applicability of the theory
to the present model. As said, we will find that the pressure fluctuation in this model are not normal, and accordingly the
stress autocorrelations are shown to decay anomalously slowly. In Sect.~\ref{summary} we provide a summary and conclusions.

\section{Materials and methods}
\label{model}

In our simulations we create amorphous granular assemblies of $N$ disks, half of which have a radius $R_1=0.35$ and the other half
with a radius $R_2=0.49$.
We focus on frictional assemblies of granular disks that are at mechanical equilibrium, having some finite pressure above the jamming point, and confined in periodic cells. To produce such meaningful granular states, we start from a dilute granular medium in which the disks are placed randomly without overlap, and progressively compress it while integrating Newton's second law with added damping, until a mechanical equilibrium is reached at a desired target pressure.

The contact forces, which include both normal and tangential components due to friction, are modeled according to the discrete element method developed by Cundall and Strack \cite{79CS}, combining a Hertzian normal force and a tangential Mindlin component. For our 2D system in the $(x,y)$ plane, consider two particles $i$ and $j$, at positions $\B r_i$, $\B r_j$ with velocities $\B v_i$, $\B v_j$ and angular velocities $\omega_i\vec e_z$, $\omega_j\,\vec e_z$. They interact only if forming a contact, i.e. if the relative normal compression $\Delta_{ij}^{(n)}=D_{ij}-r_{ij}>0$, where $r_{ij}=|\vec r_{ij}|$, $\B r_{ij}=\vec r_i-\vec r_j$, $D_{ij}=R_i+R_j$, and $R_i$, $R_j$ the radii of grains $i$ and $j$. We denote $\vec n_{ij}=\vec r_{ij}/r_{ij}$ the normal unit vector, and $\vec t_{ij}$, its transform by the $\pi/2$ rotation. The Cundall-Strack forces also depend on the elastic tangential displacement $\Delta_{ij}^{(t)}$, which is set to zero when any contact is first made and integrated numerically as long as it is maintained, using~\cite{01SEGHLP}
\begin{equation}
\frac{d \Delta_{ij}^{(t)}}{d t}= \B {v}_{ij}\cdot \vec t_{ij}-\frac{1}{2}\,(\omega_i+\omega_j) r_{ij}
\label{rotation}
\end{equation}
where $\B {v}_{ij} = \B {v}_{i} - \B {v}_{j}$. It is useful to introduce the normal and tangential component of the relative velocity at contact:
   \begin{equation}
\begin{split}
     {\B v}^{(n)}_{ij}&= ({\B v}_{ij} .\B n_{ij})\,\B n_{ij}  \\
     {\B v}^{(t)}_{ij}&= ({\B v}_{ij} .\B t_{ij})\,\B t_{ij} - \frac{1}{2}(\B \omega_i + \B \omega_j)\times \B r_{ij}.
\end{split}
\end{equation}
with $\times$ the cross product.

The Cundall-Strack force exerted by grain $j$ on $i$ is
\begin{equation}\label{eq:forces}
\begin{split}
\B F^{(n)}_{ij}&=k_n\Delta_{ij}^{(n)}\B n_{ij}-\frac{\gamma_n}{2} \B {v}^{(n)}_{ij}\\
\B F^{(t)}_{ij}&=-k_t \Delta_{ij}^{(t)}\B t_{ij}-\frac{\gamma_t}{2} \B {v}^{(t)}_{ij}
\end{split}
\end{equation}
where
\begin{equation}
\begin{split}
k_n &= k_n'\sqrt{ \Delta_{ij} R_{ij}} \ , \quad
k_t = k_t'\sqrt{ \Delta_{ij} R_{ij}} \\
\gamma_{n} &= \gamma_{n}^{'}  \sqrt{ \Delta_{ij} R_{ij}}\ , \quad
\gamma_{t} = \gamma_{t}^{'}  \sqrt{ \Delta_{ij} R_{ij}} \ .
\end{split}
\end{equation}
with $R_{ij}^{-1}\equiv R_i^{-1}+R_j^{-1}$, $k_n^{'}$ and $k_t^{'}$ the normal and tangential (resp.) spring stiffness, and $\gamma_n^{'}$ and $\gamma_t^{'}$ the viscoelastic damping constants.  The above expression for the tangential force holds only so long at it does not exceed the limit set by the Coulomb limit
\begin{equation}
  \left|F^{(t)}_{ij}\right| \le \mu F^{(n)}_{ij} \ , \label{Coulomb}
\end{equation}
where $\mu$ is a material dependent coefficient. The attainment of this limit is achieved below in two different ways. We will refer to the first as
model A: when this limit is exceeded $F^{(t)}_{ij}$ is set to $\pm\mu F^{(n)}_{ij}$; the contact slips in a dissipative fashion. In model B the limit
is achieved smoothly, with two derivatives. Following Refs.~\cite{19CGPP,19CGPPa,20BCGPZ} we choose:
\begin{equation}
\!\!\B F_{ij}^{(t)}\! =\! -k_t\delta_{ij}^{1/2}\!\left[1\!+\!\frac{t_{ij}}{t^*_{ij}} \!-\!\left(\frac{t_{ij}}{t^*_{ij}}\right)^2\right]\!t_{ij} \!\hat t_{ij} \ , ~
t^*_{ij}\! \equiv\! \mu \frac{k_n}{k_t} \delta_{ij} \ .
\label{Ft}
\end{equation}
Now the derivative of the force with respect to $t_{ij}$ vanishes smoothly at $t_{ij}=t^*_{ij}$ and Eq.~(\ref{Coulomb}) is fulfilled. In both models the limit of frictionless particles is reached when $\mu=0$.

In the present simulations we use stiffnesses $k_n=k_t=2\times 10^6$. The mass of each disk is $m=1$, and we will use it as our unit of mass. The unit of length will be $2R_1$ and time in units of $1/\sqrt{k_n}$. The friction coefficient will vary and will be reported below explicitly.
Most of our results are reported for $\mu=1$.

Simulations are performed using the open source codes, LAMMPS \cite{95Pli} and LIGGGHTS \cite{12KGHAP} to properly keep track of both the normal and the history-dependent tangential force. Initially, the grains are placed randomly in a large two dimensional box while forbidding the existence of overlaps or contacts. The system is then isotropically compressed along $x$ and $y$ directions while integrating Newton's second law with total forces and (scalar) torques on particle $i$ given by
\begin{equation}\label{eq:total}
  \begin{split}
  \B F_{i}&= \sum_{j}\B F^{(n)}_{ij} + \B F^{(t)}_{ij}  \\
  \tau_{i}&= \sum_{j}\tau_{ij}
  \end{split}
\end{equation}
with
\begin{equation}
  \tau_{ij}\equiv-\frac{1}{2}\left({\vec r}_{ij} \times\B F^{(t)}_{ij}\right)\cdot{\vec e}_z
\end{equation}
the torque exerted by $j$ onto $i$.
  In one compression step we reduce the system's area isotropically, for $10^5$ MD steps with rate (per MD step) $5\times 10^{-8}$. After each compression step, the system is allowed to relax for $5\times10^5$ MD steps so that it reaches  mechanical equilibrium. We repeat these compression and relaxation steps until the system attains a jammed (mechanically balanced) configuration at the chosen pressure. The cell is kept square
  throughout the process, and in the simulations reported below
 $L_x=L_y\simeq 106$.
Of course, in the final \emph{mechanically equilibrated states} obtained at the end of compression the total force and torque [Eq.~(\ref{eq:total})] acting on each grain vanish as well as all velocities.

\section{Theory: stress correlation in frictional granular assemblies}
\label{theory}

\subsection{Stress fields}\label{sec:stress}
The coarse-grained stress tensor $\tensor\sigma(\vec r)$ of such a system reads~\cite{02GG}:
\begin{equation}\label{eq:gg}
{\sigma}_{\alpha\beta}(\vec r)=-\frac{1}{2}\sum_{i,j; i\ne j} F_{ij}^\alpha r_{ij}^\beta\int_0^1\d s\,\phi(\vec r-\vec r_i+s\vec r_{ij})
\end{equation}
where $\alpha$, $\beta$ refer to Cartesians coordinates, and $\phi$ is the coarse-graining function, which integrates (in 2D) to unity and vanishes beyond a cut-off $r_c$.
This expression is nothing but the convolution by $\phi$ of Hardy's microscopic stress~\cite{90ECM} $\tensor\sigma^\delta$ which, in Fourier space, reads:
\begin{equation}\label{eq:hardy}
\widehat{\sigma}_{\alpha\beta {\vec k}}^{\delta}=\frac{1}{2A}\,\sum_{i,j,i\ne j}\,F_{ij}^{\alpha}r_{ij}^{\beta}\frac{e^{-i{\vec k}\cdot {{\vec r}_i}}-e^{-i{\vec k}\cdot {{\vec r}_j}}}{i\vec k\cdot {\vec r}_{ij}}
\end{equation}
As usual, we use hats to denote Fourier transforms. Our cell being periodic, the above function is defined for all $k_\alpha=\frac{2\pi}{L_\alpha}n_\alpha$, with $\vec n=(n_x,n_y)$ a pair of integers.

We immediately note that the above-defined stress is not tensor-symmetric. Indeed its antisymmetric component is
\begin{equation}
\widehat{\sigma}_{xy {\vec k}}^{\delta}-\widehat{\sigma}_{yx {\vec k}}^{\delta}=\frac{1}{A}\,\sum_{i,j,i\ne j}\,\tau_{ij}\frac{e^{-i{\vec k}\cdot {{\vec r}_i}}-e^{-i{\vec k}\cdot {{\vec r}_j}}}{i\vec k\cdot {\vec r}_{ij}}
\label{stressrelation}
\end{equation}
where $\tau_{ij}$, the torque exerted by grain $j$ onto grain $i$, is non-zero in general.  We note that the resulting torque on any grain $\tau_i=\sum_{j}\tau_{ij}=0$ at equilibrium, although the above expression cannot be reorganized to separate the $\tau_i$'s. This parallels the fact that stress is non-zero at mechanical equilibrium even though the resulting force on each grain vanishes.

Let us check that our stress fields are divergence-free in mechanically balanced states. The divergence of stress is the vector field $i k^\beta\widehat{\sigma}_{\alpha\beta {\vec k}}^{\delta}$ (we use the convention of implicit summation on repeated indices). From~(\ref{eq:hardy}), we immediately obtain:
\begin{equation}\label{eq:div}
i k^\beta\widehat{\sigma}_{\alpha\beta {\vec k}}^{\delta}=\frac{1}{A}\,\sum_{i}\,e^{-i{\vec k}\cdot {{\vec r}_i}}\,F_{i}^{\alpha}
\end{equation}
which shows the desired result since $F_{i}^{\alpha}=0$.

The key question we address here is what is the nature of stress correlations in mechanically balanced states when the antisymmetric part of stress is non-zero. Following Ref.~\cite{17Lem}, we introduce a vector representation for stress based on the notion of \emph{spherical tensors}. Since here stress is non-symmetric, this representation must comprise four \emph{spherical} components, which we define as follows:
\begin{equation}
\label{eq:stress}
\begin{split}
\sigma_1&=-\frac{1}{2}\,\left(\sigma_{xx}+\sigma_{yy}\right)\\
\sigma_2&=\hphantom{-}\frac{1}{2}\,\left(\sigma_{xx}-\sigma_{yy}\right)\\
\sigma_3&=\hphantom{-}\frac{1}{2}\,\left(\sigma_{xy}+\sigma_{yx}\right)\\
\sigma_4&=\hphantom{-}\frac{1}{2}\,\left(\sigma_{xy}-\sigma_{yx}\right)
\end{split}
\end{equation}
It will be useful to treat the set of these four ``Cartesian'' components as the vector $\utilde\sigma=(\sigma_1,\sigma_2,\sigma_3,\sigma_4)$.
\begin{figure}
\includegraphics[scale=0.20]{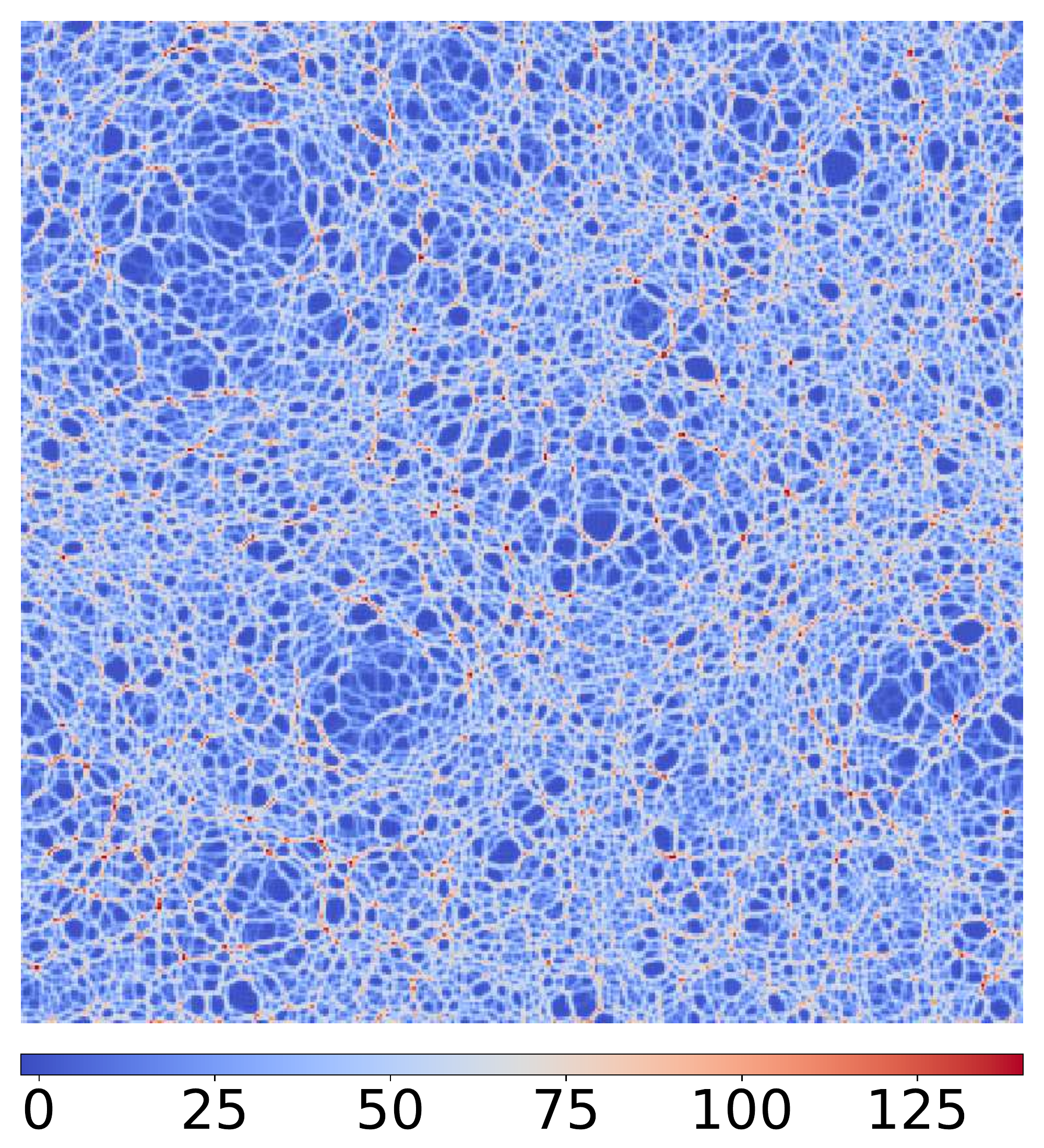}
\includegraphics[scale=0.20]{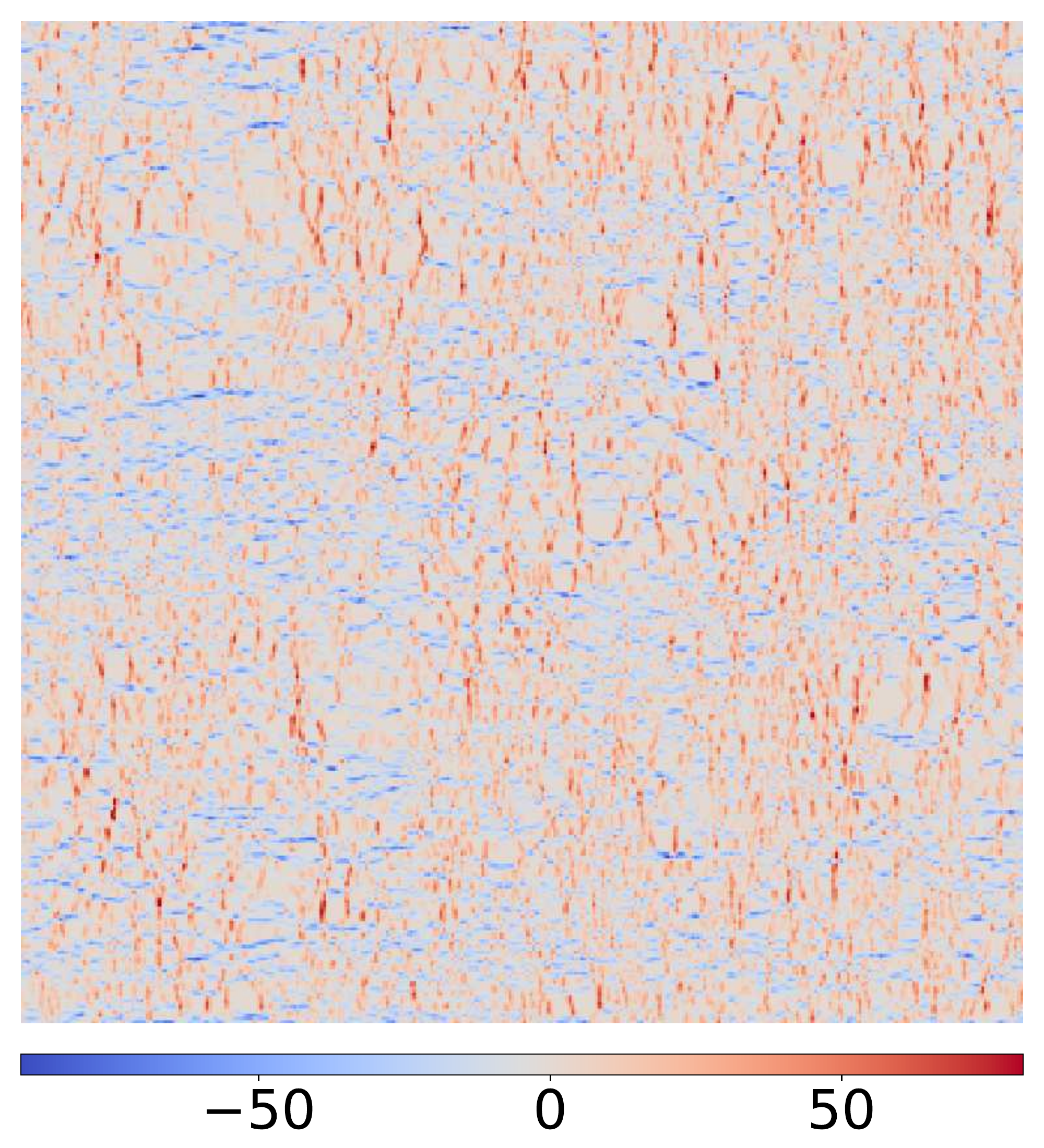}
\includegraphics[scale=0.20]{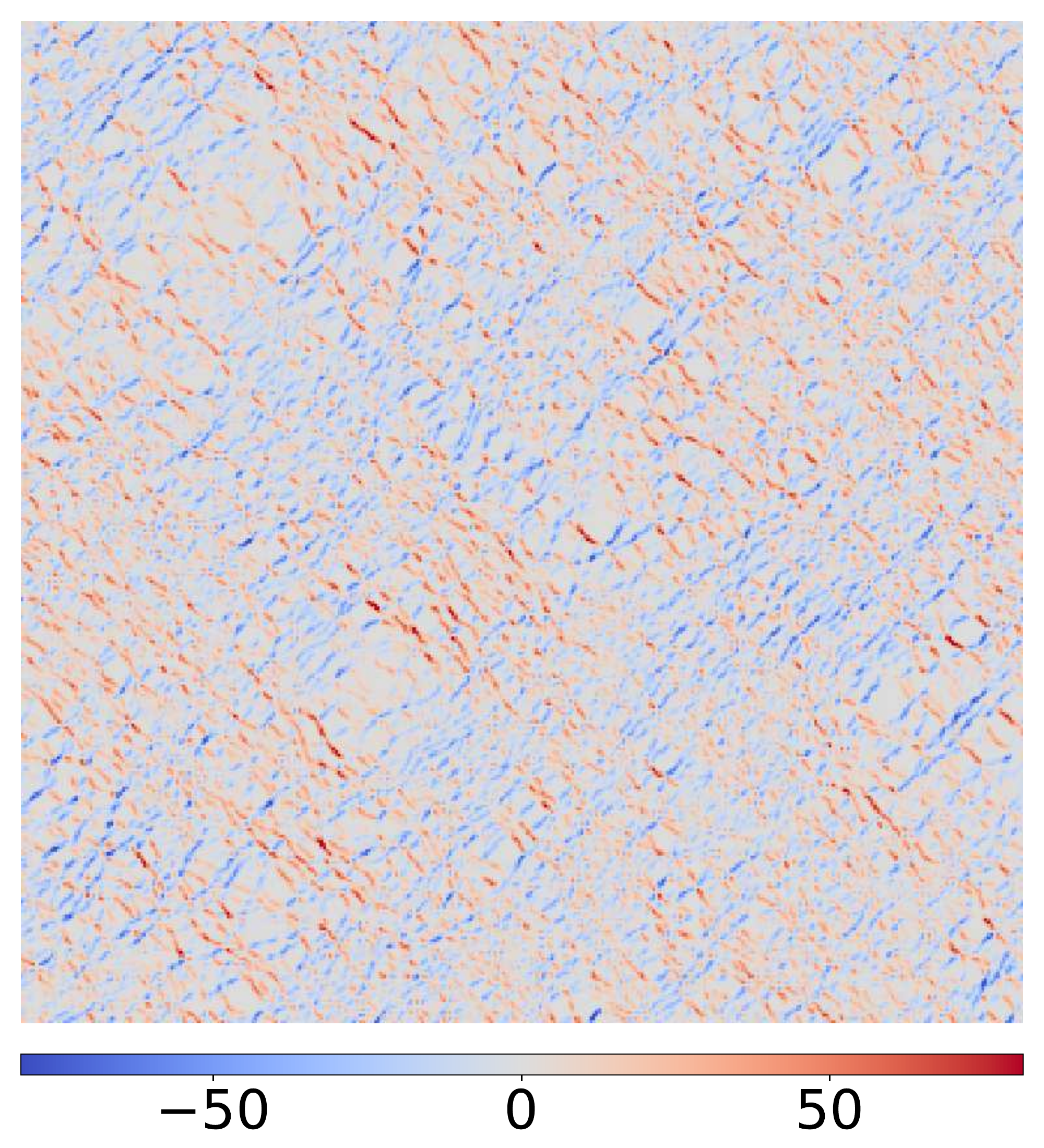}
\includegraphics[scale=0.20]{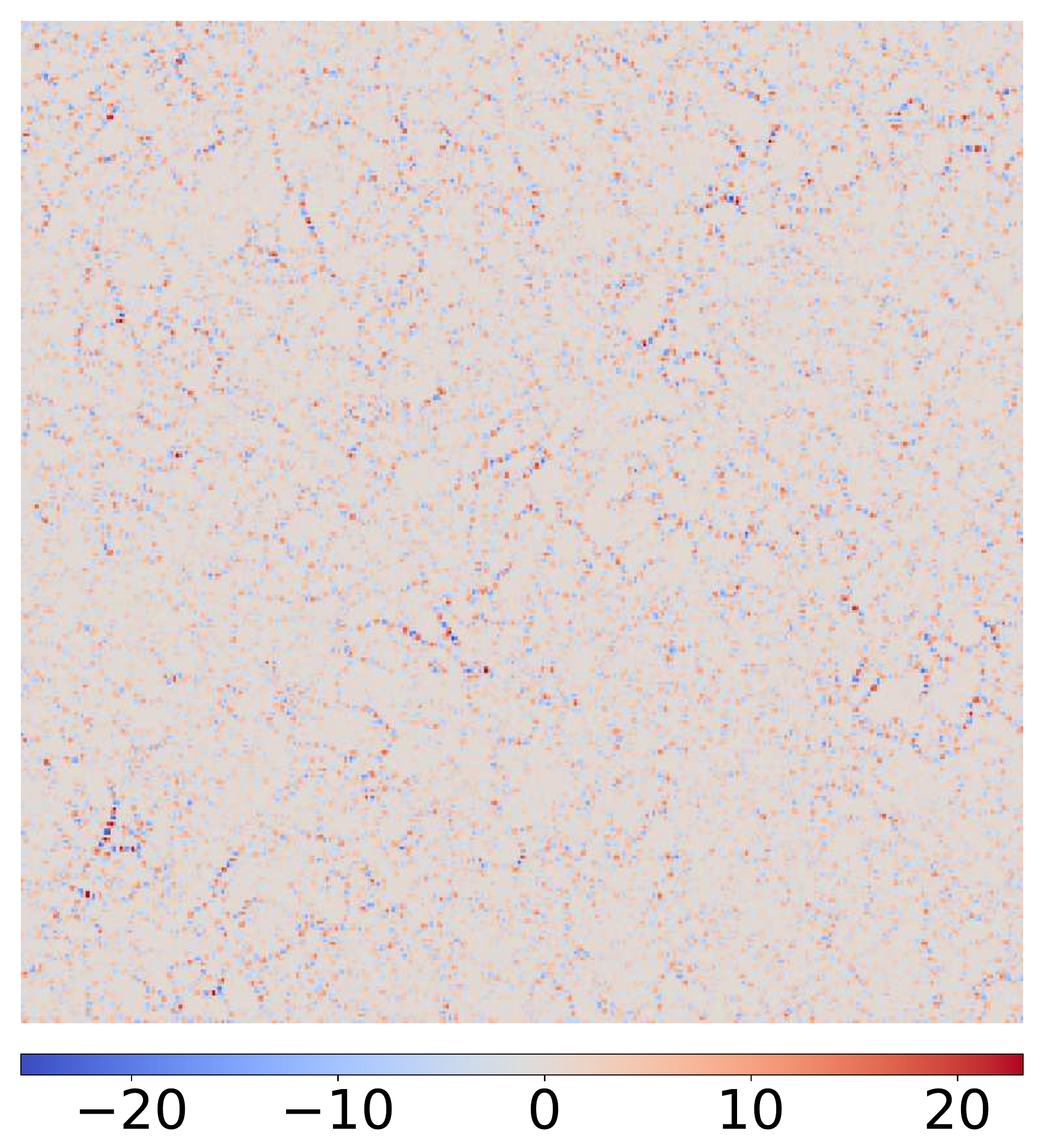}
\caption{Coarse-grained fields in real space. From top to bottom and left to right we show $\sigma_1$, then $\sigma_2$, $\sigma_3$
and $\sigma_4$, see the definition in Eq.~(\ref{eq:stress}). Note that $\sigma_1$ and $\sigma_4$ are isotropic,
$\sigma_2$ displays orientation along the $x$ and $y$ axes, while $\sigma_3$ is oriented along the diagonals. }
\label{realspace}
\end{figure}
Typical values of our four fields are represented on Fig.~\ref{realspace}, as computed using the coarse-graining function $\phi(\vec r)=\frac{15}{8\pi r_c^2}(1-(r/r_c)^4)^2$ for $r<r_c$, $\phi(\vec r)=0$ for $r>r_c$ . We see that, as in previous studies the pressure $\sigma_1$ is isotropic, while the two deviatoric stresses are clearly anisotropic and present patterns clearly suggestive of long-range correlations. However, in contrast with previous works~\cite{14Lem,17Lem}, the tensor-asymmetry $\sigma_4$, although of smaller amplitude than the other fields, is non-zero; it also does not appear to present any evident anisotropy.

\subsection{Stress autocorrelations}

From now on, we will work only with Hardy's stress and thus will drop the $\delta$ indices to simplify our notation. Moreover, our analysis will proceed in Fourier space, where the Cartesian components of stress $\utilde{\widehat{\sigma}}=(\widehat\sigma_1,\widehat\sigma_2,\widehat\sigma_3,\widehat\sigma_4)$ are defined just as in Eq.~(\ref{eq:stress}). In our translation-invariant systems, the autocorrelation matrix of these Cartesian spherical stress components is:
\begin{equation}
\label{fourcarcor}
\uttilde{\widehat{C}}_{\vec k}=\frac{1}{A}\,\left\langle\utilde{\widehat{\sigma}}_{\vec k}\,\utilde{\widehat{\sigma}}_{\vec k}^*\right\rangle_c
\end{equation}
with ${}^*$ the complex conjugate. Here, juxtaposition is used to denote the tensor product and $\langle AB \rangle_c=\langle AB\rangle-\langle A\rangle\langle B \rangle$ the second cumulant for the ensemble average.

\begin{figure}
\includegraphics[scale=0.35]{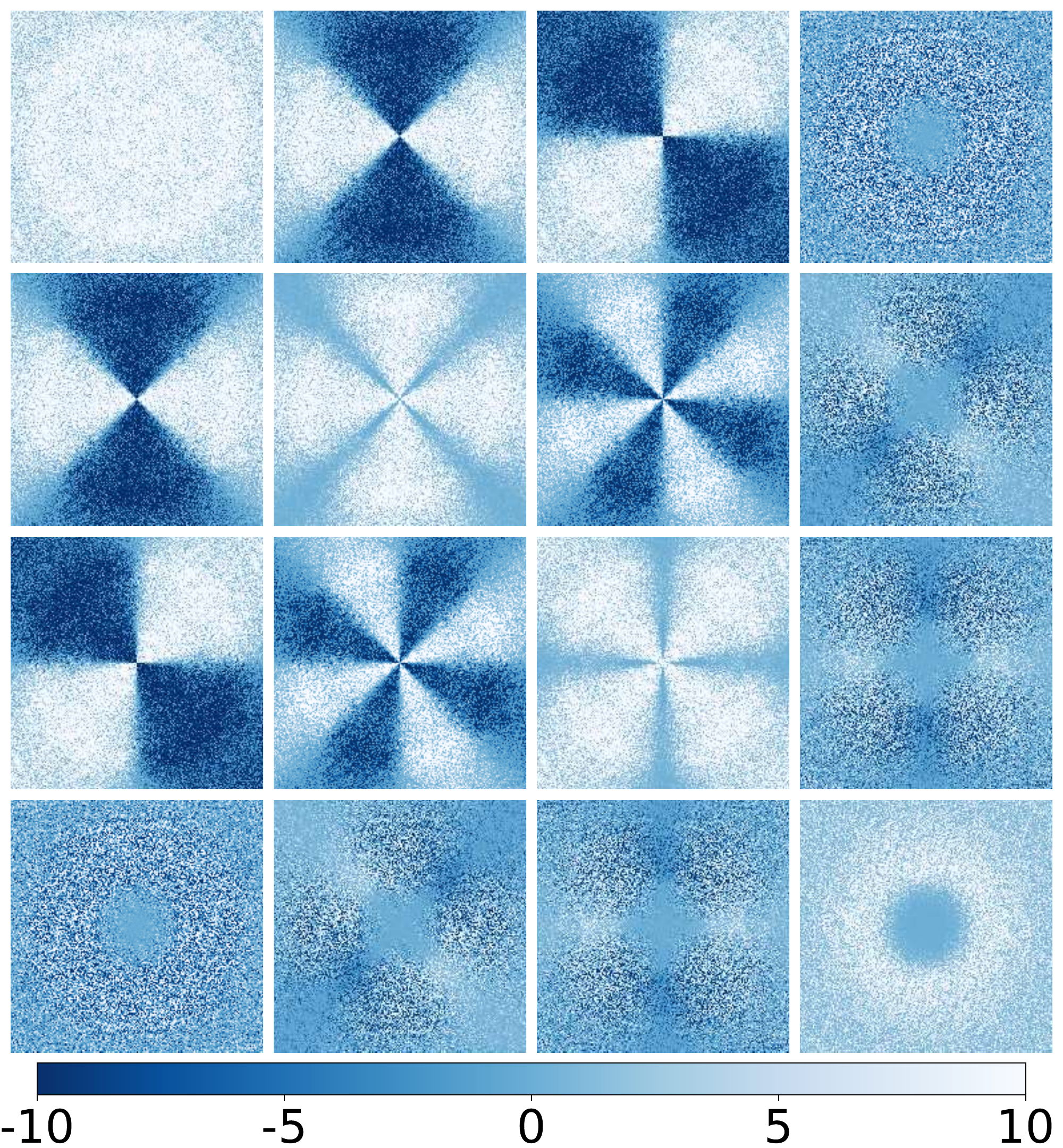}
\caption{The real-valued fields ${{\widehat{C}}}_{\vec k\,ab}$ displayed as a (symmetric) matrix.
In each frame, the origin is placed at the center. The order is such that the first row exhibits the 1,1; 1,2; 1,3 and 1,4 components of the matrix Eq.~\ref{fourcarcor}, the second row starts
with 2,1 etc.  To be able to distinguish the features of
all fields on the same color scale we have multiplied $\widehat\sigma_4$ by a factor of three. }
\label{fig:C}
\end{figure}
For the sake of illustration, we report in Fig.~\ref{fig:C} the components of $\uttilde{\widehat{C}}_{\vec k}$ as a matrix of fields. Anisotropies are clearly seen, not only in the autocorrelations ${\widehat{C}}_{22,\,\vec k}$ and ${\widehat{C}}_{33,\,\vec k}$, where they are expected, but also in all the rest except in the pressure (${\widehat{C}}_{11,\,\vec k}$) and torque (${\widehat{C}}_{44,\,\vec k}$) density autocorrelation and their cross-correlations which appear
to vanish. The submatrix $\widehat{C}_{ab}$ with $a,b=1,\ldots,3$ presents the same symmetries as in previous works~\cite{17Lem}, but the existence of anisotropic correlations between the torque density and other fields is unexpected.

Let us now consider the stress vector components in the basis $(\vec e_k,\vec e_\phi)$ of cylindrical coordinates for an arbitrary non-zero wavevector $\vec k$:
\begin{equation}
  \label{eq:stress:radial}
\begin{split}
\widehat{\sigma}^{\uvec k}_{1\,\vec k}&=-\frac{1}{2}\left( \widehat{\sigma}_{kk\,\vec k}+ \widehat{\sigma}_{\phi\phi\,\vec k}\right)\\
\widehat{\sigma}^{\uvec k}_{2\,\vec k}&=\hphantom{-}\frac{1}{2}\left( \widehat{\sigma}_{kk\,\vec k}- \widehat{\sigma}_{\phi\phi\,\vec k}\right)\\
\widehat{\sigma}^{\uvec k}_{3\,\vec k}&=\hphantom{-}\frac{1}{2}\,\left(\widehat{\sigma}_{k\phi\,\vec k}+\widehat{\sigma}_{\phi k\,\vec k}\right)\\
\widehat{\sigma}^{\uvec k}_{4\,\vec k}&=\hphantom{-}\frac{1}{2}\,\left(\widehat{\sigma}_{k\phi\,\vec k}-\widehat{\sigma}_{\phi k\,\vec k}\right)
\end{split}
\end{equation}
where $\uvec k\equiv \vec k/k$ denotes the considered direction in reciprocal space. As before, these \emph{radial} components define a vector, denoted $\utilde{\widehat{\sigma}}^{\uvec k}_{\vec k}=(\widehat{\sigma}_{1\,\vec k}^{\uvec k},\widehat{\sigma}_{2\,\vec k}^{\uvec k},\widehat{\sigma}_{3\,\vec k}^{\uvec k},\widehat{\sigma}_{4\,\vec k}^{\uvec k})$.
To understand the role of material isotropy, we introduce the autocorrelation matrix of these radial components, $\uttilde{\mathring{\widehat{C}}}_{\vec k}$ which, at any $\vec k$, is:
\begin{equation}
\uttilde{\mathring{\widehat{C}}}_{\vec k}=\frac{1}{A}\langle \utilde{\widehat{\sigma}}_{{\vec k}}^{\uvec k}(\utilde{\widehat{\sigma}}_{{\vec k}}^{\uvec k})^*\rangle_c
\end{equation}
We call this object the "radial spherical" autocorrelation matrix.

The advantage of our vector representations of stress is that it permits to deal with rotation transforms of stress using quite simple relations~\cite{17Lem}. Indeed, the above defined Cartesian ($\utilde{\widehat{\sigma}}_{\vec k}$) and radial ($\utilde{\widehat{\sigma}}^{\uvec k}_{\vec k}$) vectors are related by the simple expression:
\begin{equation}\label{eq:cstors}
  \utilde{\widehat{\sigma}}^{\uvec k}_{\vec k}=\mathbcal{D}^{\uvec k}\cdot\utilde{\widehat{\sigma}}_{\vec k}
\end{equation}
with
\begin{equation}
  \label{eq:Rr}
\mathbcal{D}^{\uvec k}=
\left(
  \begin{matrix}
    1&0&0&0\\
    0&\quad\cos2\phi&\quad\sin2\phi&0\\
    0&-\sin2\phi&\quad\cos2\phi&0\\
    0&0&0&1\\
  \end{matrix}
\right)
\end{equation}
It follows that the Cartesian spherical and radial spherical autocorrelation matrices verify:
\begin{equation}
\label{eq:fourier:radial:corr}
\uttilde{\mathring{\widehat{C}}}_{\vec k} =\mathbcal{D}^{\uvec k}\cdot\uttilde{\widehat{C}}_{\vec k}\cdot(\mathbcal{D}^{\uvec k})^T
\end{equation}

\subsection{Material isotropy}

Let us now examine the consequences of material symmetries on stress correlations. First, we note that our jammed ensembles verify by construction spatial inversion symmetry, which entails that both $\uttilde{\widehat{C}}_{\vec k}$ and $\uttilde{\mathring{\widehat{C}}}_{\vec k}$ are real-valued and spatially symmetric in the senses
that the (i,j) and (j,i) components present the same spatial dependence. They are indeed spatially symmetric as we observed for $\uttilde{{\widehat{C}}}_{\vec k}$ in Fig.~\ref{fig:C}.

Material isotropy is not expected to hold at all distances in finite size systems, due to periodic boundary conditions. But it must arise in the infinite size limit, and should hence progressively be achieved at any fixed $\vec k$ when $L\to\infty$. To discuss material isotropy, we are thus led to consider the infinite medium stress autocorrelations $\uttilde{\mathring{\widehat{C}}}^\infty(\vec k)$ and $\uttilde{{\widehat{C}}}^\infty(\vec k)$, which are continuous functions of $\vec k$.

Material isotropy means that the infinite medium ensemble is invariant under rotations, but also under reflections, i.e. under all unitary transformations. It is important to realize that, in 2D, the point reflection has a determinant $=1$; therefore, we do not exhaust all unitary transformations of the medium by only considering point inversion symmetry (as we have already done above) and rotations. We will need to explicitly take into account axial reflection invariance.

Proper rotation invariance amounts to the property that a radially symmetric stress autocorrelation is independent of direction $\uvec k$, i.e. is a function of the amplitude $k$ only:
\begin{equation}\label{eq:rotation}
  \uttilde{\mathring{\widehat{C}}}^\infty({\vec k}) = \uttilde{\mathring{\widehat{C}}}^\infty({k}) = \uttilde{\widehat{C}}^\infty({k\vec e_x})
\end{equation}
where the last equality corresponds to the specific case when $\vec k=k\,\vec e_x$, i.e. $\theta=0$. This equation makes it obvious that the
Cartesian symmetric autocorrelation, and hence Cartesian stress fields, should present spatial anisotropies. Indeed, inverting Eq.~(\ref{eq:fourier:radial:corr}) we now have:
\begin{equation}\label{eq:RS2CS}
\uttilde{\widehat{C}}^\infty({\vec k})=(\mathbcal{D}^{\uvec k})^T\cdot\uttilde{\mathring{\widehat{C}}}^\infty(k)\cdot\mathbcal{D}^{\uvec k}
\end{equation}
which demonstrates that, since $\uttilde{\mathring{\widehat{C}}}^\infty$ is spatially isotropic, $\uttilde{\widehat{C}}^\infty({\vec k})$ is not, but presents trivial anisotropies originating from the right and left products with rotation matrices.

To guarantee material isotropy, we are now left with requiring reflection symmetry about one chosen axis. The invariance of $\uttilde{\mathring{\widehat{C}}}^\infty$ about axis $\uvec k$ is equivalent to that of $\uttilde{\widehat{C}}^\infty$ about the $x$ axis, i.e. under the $y\to-y$ transformation, which acts on stress as:
\begin{equation}
  \utilde\sigma\to\mathbcal{D}_{-1}\cdot\utilde\sigma
\end{equation}
with
\begin{equation}
  \label{eq:axial}
\mathbcal{D}_{-1}=
\left(
  \begin{matrix}
    1&0&0&0\\
    0&1&0&0\\
    0&0&-1&0\\
    0&0&0&-1\\
  \end{matrix}
\right)
\end{equation}
It follows that reflection-invariance amounts to requiring that the radially symmetric autocorrelations satisfy:
\begin{equation}
\uttilde{\mathring{\widehat{C}}}^\infty=\mathbcal{D}_{-1}\cdot\uttilde{\mathring{\widehat{C}}}^\infty\cdot\mathbcal{D}_{-1}^T
\end{equation}
Group theory (Schur's first lemma) then demonstrates that $\uttilde{\mathring{\widehat{C}}}^\infty$ verifies this property iff it is of the block form:
\begin{equation}
  \label{axial2}
\uttilde{\mathring{\widehat{C}}}^\infty=
\left(
  \begin{matrix}
    \mathring{\widehat{C}}_1&\mathring{\widehat{C}}_2&0&0\\
    \mathring{\widehat{C}}_2&\mathring{\widehat{C}}_3&0&0\\
    0&0&\mathring{\widehat{C}}_4&\mathring{\widehat{C}}_5\\
    0&0&\mathring{\widehat{C}}_5&\mathring{\widehat{C}}_6\\
  \end{matrix}
\right)
\end{equation}
since we already know that $\uttilde{\mathring{\widehat{C}}}^\infty$ is a symmetric matrix. The above expression only involves six spatially isotropic functions $\mathring{\widehat{C}}_a(k)$, with $a=1,\ldots,6$.

Note that the arguments we have developed here in Fourier space can be carried out identically in real space, and entail that the radially symmetric autocorrelations $\uttilde{\mathring{{C}}}^\infty$ present the same form, fully determined by six spatially isotropic functions $\mathring{{C}}_a(r)$, $a=1,\ldots,6$.

\subsection{Mechanical balance}

We checked in Sec.~\ref{sec:stress} that coarse-grained Hardy's stress fields are, as expected, strictly divergence-free in mechanically balanced (jammed) states. Mechanical balance thus reads
\begin{equation}
i k^\beta\widehat{\sigma}_{\alpha\beta {\vec{k}}}=0
\end{equation}
which is easily recast in the radial frame, as:
\begin{equation}
\forall\vec k\ne\vec 0\,,\qquad\widehat{\sigma}_{kk}=
\widehat{\sigma}_{\phi k}=0
\end{equation}
In terms of vector components, in view of Eq.~(\ref{eq:stress:radial}), it becomes:
\begin{equation}\label{eq:mech}
\forall\vec k\ne\vec 0\,,\qquad \widehat{\sigma}^{\uvec k}_{1\,\vec k}=\widehat{\sigma}^{\uvec k}_{2\,\vec k}
  \quad{\rm and}\quad
  \widehat{\sigma}^{\uvec k}_{3\,\vec k}=\widehat{\sigma}^{\uvec k}_{4\,\vec k}
\end{equation}

We are interested in systems that are both mechanically balanced and materially isotropic. Plugging Eq.~(\ref{eq:mech}) into~(\ref{axial2}), we now see that the radially symmetric autocorrelation matrix must then be of the form:
\begin{equation}
  \label{axial3}
\uttilde{\mathring{\widehat{C}}}^\infty=
\left(
  \begin{matrix}
    \mathring{\widehat{C}}&\mathring{\widehat{C}}&0&0\\
    \mathring{\widehat{C}}&\mathring{\widehat{C}}&0&0\\
    0&0&\mathring{\widehat{C}}'&\mathring{\widehat{C}}'\\
    0&0&\mathring{\widehat{C}}'&\mathring{\widehat{C}}'\\
  \end{matrix}
\right)
\end{equation}
which now involve just two spatially isotropic functions $\mathring{\widehat{C}}(k)$ and $\mathring{\widehat{C}}'(k)$ which we will identify shortly. The matrix structure we have obtained here differs from that found in previous works~\cite{14Lem,17Lem}, which only involved the pressure autocorrelation $\mathring{\widehat{C}}$ since stress was symmetric and hence $\mathring{\widehat{C}}'\equiv0$.

The relative simplicity of the above expression permits us to use~(\ref{eq:RS2CS}) and obtain a general expression for the Cartesian symmetric autocorrelations:
\begin{widetext}
\begin{equation}
\uttilde{{\widehat{C}}}^\infty=
\left(
  \begin{matrix}
    \mathring{\widehat{C}}&\cos2\phi\,\mathring{\widehat{C}}&\sin2\phi\,\mathring{\widehat{C}}&0\\
    \cos2\phi\,\mathring{\widehat{C}}&\frac{1}{2}\left(\mathring{\widehat{C}}+\mathring{\widehat{C}}'\right)+\frac{1}{2}\cos4\phi\left(\mathring{\widehat{C}}-\mathring{\widehat{C}}'\right)&\frac{1}{2}\sin4\phi\left(\mathring{\widehat{C}}-\mathring{\widehat{C}}'\right)&-\sin2\phi\,\mathring{\widehat{C}}'\\
    \sin2\phi\,\mathring{\widehat{C}}&\frac{1}{2}\sin4\phi\left(\mathring{\widehat{C}}-\mathring{\widehat{C}}'\right)&\frac{1}{2}\left(\mathring{\widehat{C}}+\mathring{\widehat{C}}'\right)-\frac{1}{2}\cos4\phi\left(\mathring{\widehat{C}}-\mathring{\widehat{C}}'\right)&\cos2\phi\,\mathring{\widehat{C}}'\\
    0&-\sin2\phi\,\mathring{\widehat{C}}'&\cos2\phi\,\mathring{\widehat{C}}'&\mathring{\widehat{C}}'\\
  \end{matrix}
\right)
\end{equation}
Having in mind Eqs.~(\ref{stressrelation}) and (\ref{eq:stress})  we see very clearly that the functions $\mathring{\widehat{C}}$ and $\mathring{\widehat{C}}'$ are respectively the autocorrelations of local pressure and local torque density.

The real-space stress autocorrelation in the inverse Fourier transform of this expression:
\begin{equation}
{\uttilde{{C}}}^\infty({\vec r})=\frac{1}{(2\pi)^2}\,\int\d\vec k\,e^{i\vec k\cdot\vec r}\,\uttilde{\widehat{C}}^\infty({\vec k})
\end{equation}
To perform its calculation explicitly, we use:
\begin{equation}
\begin{split}
\int\d\vec k\,e^{i\vec k\cdot\vec r}\,\widehat{f}(k)\cos(m\phi)
&=2\pi i^m \cos(m\theta)\int_0^\infty\d k\,k\widehat{f}(k)J_m(kr)\\
\int\d\vec k\,e^{i\vec k\cdot\vec r}\,\widehat{f}(k)\sin(m\phi)
&=2\pi i^m \sin(m\theta)\int_0^\infty\d k\,k\widehat{f}(k)J_m(kr)
\end{split}
\end{equation}
with $J_m$ the Bessel function of the first kind of order $m$.
It then immediately appears that the real-space stress autocorrelation is of the form:
\begin{equation}\label{eq:C:back}
\qquad{\uttilde{{C}}}^\infty({\vec r})=
\left(
  \begin{matrix}
    \mathring{\widehat{C}}^{(0)}&-\cos2\theta\,\mathring{\widehat{C}}^{(2)}&-\sin2\theta\,\mathring{\widehat{C}}^{(2)}&0\\
    -\cos2\theta\,\mathring{\widehat{C}}^{(2)}&\frac{\mathring{\widehat{C}}^{(0)}+\mathring{\widehat{C}}'^{(0)}}{2}+\cos4\theta\ \frac{\mathring{\widehat{C}}^{(4)}-\mathring{\widehat{C}}'^{(4)}}{2}&\sin4\theta\ \frac{\mathring{\widehat{C}}^{(4)}-\mathring{\widehat{C}}'^{(4)}}{2}&\sin2\theta\,\mathring{\widehat{C}}'^{(2)}\\
    -\sin2\theta\,\mathring{\widehat{C}}^{(2)}&\sin4\theta\ \frac{\mathring{\widehat{C}}^{(4)}-\mathring{\widehat{C}}'^{(4)}}{2}&\frac{\mathring{\widehat{C}}^{(0)}+\mathring{\widehat{C}}'^{(0)}}{2}-\cos4\theta\ \frac{\mathring{\widehat{C}}^{(4)}-\mathring{\widehat{C}}'^{(4)}}{2}&-\cos2\theta\,\mathring{\widehat{C}}'^{(2)}\\
    0&\sin2\theta\,\mathring{\widehat{C}}'^{(2)}&-\cos2\theta\,\mathring{\widehat{C}}'^{(2)}&\mathring{\widehat{C}}'^{(0)}\\
  \end{matrix}
\right)
\end{equation}
\end{widetext}
where for any spatially isotropic function $\widehat{f}(k)$:
\begin{equation}\label{eq:transforms}
\widehat{f}^{(m)}(r)\equiv\frac{1}{2\pi}\,\int_0^\infty\d k\,k\,\widehat{f}(k)\,J_m(kr)\ .
\end{equation}
This expression defines, for each $m$, a scalar function $\widehat{f}^{(m)}(r)$ as a functional transform of the scalar function $\widehat{f}(k)$.
To assess that these transforms are well-defined note, following~\cite{18Lem}, that any scalar function of $k$ or $r$ (resp.) can be seen as defining a spatially isotropic function in Fourier or real (resp.) space. Moreover, in an arbitrary dimension $d$ the inverse Fourier transform of any spatially isotropic function $\widehat{f}(k)$ is:
\begin{equation}
f(r)=(2\pi)^{-d/2}\,\int_0^\infty\d k\ \frac{k^{d/2}}{r^{d/2-1}}\,\widehat{f}(k)\,J_{\frac{d}{2}-1}(kr)
\end{equation}
We thus recognize in the rhs of Eq.~(\ref{eq:transforms}), up to a $k$-independent prefactor, the inverse Fourier transform $\mathcal{F}_{2m+2}^{-1}$ of the spatially isotropic function $\widehat{f}(k)/{k^m}$ in dimension $2m+2$. So, the above transform can be recast as:
\begin{equation}\label{eq:inverse:d}
\widehat{f}^{(m)}(r)=(2\pi)^m\,r^m\,\mathcal{F}_{2m+2}^{-1}\left[\frac{\widehat{f}(k)}{k^m}\right]
\end{equation}
This relation guarantees that the above-defined inverse transforms are well-defined at least in the sense of distributions. Note that the $m=0$ transforms are just the inverse 2D Fourier transforms, as expected, since $\mathring{\widehat{C}}^{(0)}$, the pressure autocorrelation, is just the inverse Fourier transform of $\mathring{\widehat{C}}$. Likewise $\mathring{\widehat{C}}'^{(0)}$ is the real space autocorrelation of the local torque density.

The associated radially symmetric form is:
\begin{widetext}
\begin{equation}\label{eq:C:back:RS}
  \qquad{\uttilde{{C}}}^\infty({\vec r})=
  \left(
    \begin{matrix}
      \mathring{\widehat{C}}^{(0)}&-\mathring{\widehat{C}}^{(2)}&0&0\\
      -\mathring{\widehat{C}}^{(2)}&\frac{1}{2}\left(\mathring{\widehat{C}}^{(0)}+\mathring{\widehat{C}}^{(4)}+\mathring{\widehat{C}}'^{(0)}-\mathring{\widehat{C}}'^{(4)}\right)&0&0\\
      0&0&\frac{1}{2}\left(\mathring{\widehat{C}}^{(0)}-\mathring{\widehat{C}}^{(4)}+\mathring{\widehat{C}}'^{(0)}+\mathring{\widehat{C}}'^{(4)}\right)&-\mathring{\widehat{C}}'^{(2)}\\
      0&0&-\mathring{\widehat{C}}'^{(2)}&\mathring{\widehat{C}}'^{(0)}\\
    \end{matrix}
  \right)
\end{equation}
\end{widetext}
This expression establishes that the $r$-dependence of the real-space autocorrelation is entirely determined by the transforms $\mathring{\widehat{C}}^{(m)}$ and $\mathring{\widehat{C}}'^{(m)}$ with $m=0,2,4$. It thus opens the way towards a rational understanding of how the low $k$ behavior of $\uttilde{\mathring{\widehat{C}}}^\infty$, i.e. of the two functions $\mathring{\widehat{C}}$ and $\mathring{\widehat{C}}'$ determine the decay with distance in real space.

If a function $\widehat{f}$ is regular at the origin then its inverse Fourier transform is a rapidly (i.e. essentially exponentially) decaying function. In other cases, let us recall that, in dimension $d$, for any $s>-d$, provided $s\ne 0, 2, 4,\ldots$, the inverse Fourier transform of $k^s$, which is rigorously defined in the sense of tempered distributions~\cite{Riesz1949,Landkof1972}, is:
\begin{equation}\label{eq:inverse:explicit}
  \mathcal{F}_{d}^{-1}\left[k^s\right]=\frac{c_{d,s}}{r^{d+s}}
\end{equation}
with the constant
\begin{equation}
c_{d,s}=\frac{2^s}{\pi^{\frac{d}{2}}}\,\frac{\Gamma\left(\frac{d+s}{2}\right)}{\Gamma\left(-\frac{s}{2}\right)}
\end{equation}
This relation applies, in particular, to all values of $s$ on the interval $0>s>-d$.

An important special case is when pressure presents normal fluctuations, that is when the fluctuations of the local, domain-averaged, pressure decay normally as the inverse averaging domain volume. In that case, indeed, the pressure autocorrelation $\mathring{\widehat{C}}$ converges in the $k\to0$ limit~\cite{18Lem}. It then appears from Eqs.~(\ref{eq:inverse:d}) and~(\ref{eq:inverse:explicit}), that: (i) the real space pressure autocorrelation, $\mathring{\widehat{C}}^{(0)}$, decays exponentially; (ii) meanwhile, $m=2$ and 4 transforms present $1/r^2$ power law decay since: $\mathring{\widehat{C}}^{(m)}(r)=(2\pi)^m\,r^m\,\mathcal{F}_{2m+2}^{-1}\left[\frac{\mathring{\widehat{C}}(k)}{k^m}\right]\sim(2\pi)^m\,r^m\,\mathcal{F}_{2m+2}^{-1}\left[\frac{\mathring{\widehat{C}}(0)}{k^m}\right]\propto1/r^2$.

The above arguments, however, are far more general, and permit to deduce the long-range spatial decay in cases when the pressure autocorrelation does not converge in the $k\to0$ limit, but scales with $k$ to a negative power. If $\mathring{\widehat{C}}\sim k^{-\nu}$ at low $k$, with $\nu<d$, we then find for all $m=0,2,4$: $\mathring{\widehat{C}}^{(m)}\sim1/r^{2-\nu}$, which decays more slowly than $1/r^2$.

\begin{figure}
	\includegraphics[scale=0.35]{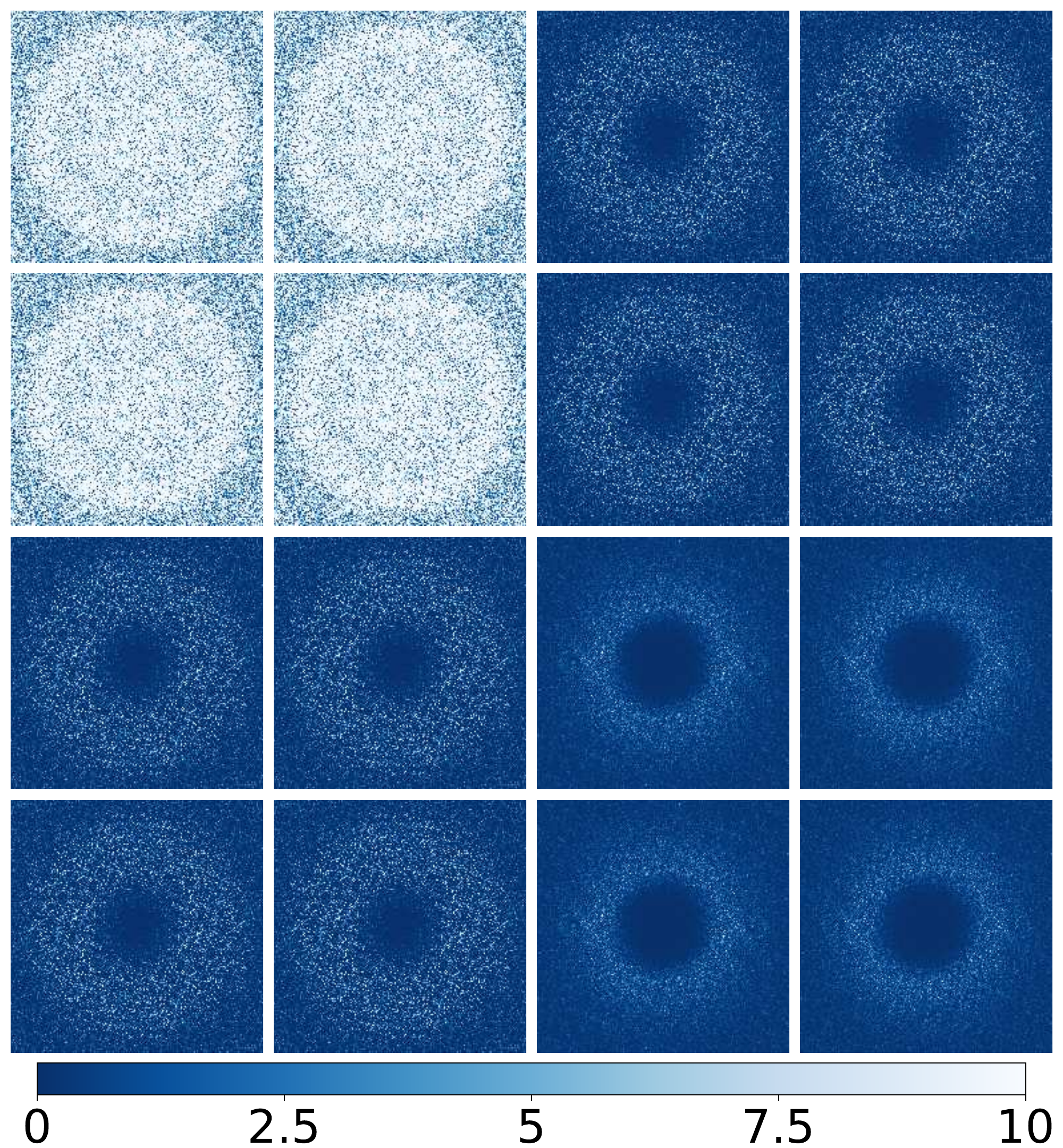}
	\caption{Real part of different components of $\tenq[2]{\mathring{\widehat{C}}}_{\vec k}$.
		The origin is placed at the center of each plot. The white speckles in the off diagonal fields indicate fluctuations around zero, and
		cf Fig.~\ref{offdia} for further evidence. }
	\label{cr}
\end{figure}


\section{Comparison of theory and simulations}
\label{numerics}

\subsection{Visual confirmation of Eq.~(\ref{axial3})}
\label{kspace}

In Fig.~\ref{cr} we plot $\tenq[2]{\mathring{\widehat{C}}}_{\vec k ab}$ of Eq.~(\ref{axial3}) vs. k for all $a,b \in {1,2}$ for the frictional system. Here we show model A, but model B results in essentially the same images. It is clear from this plot that all the fields are spatially isotropic. However we note that the off-diagonal fields which
should vanish exactly exhibit large remnant fluctuations which we will show hereafter that they result from numerical inaccuracies.
AS predicted by Eq.~(\ref{axial3}) all the four fields in each diagonal block are identical.
\begin{figure}
		 \includegraphics[scale=0.12]{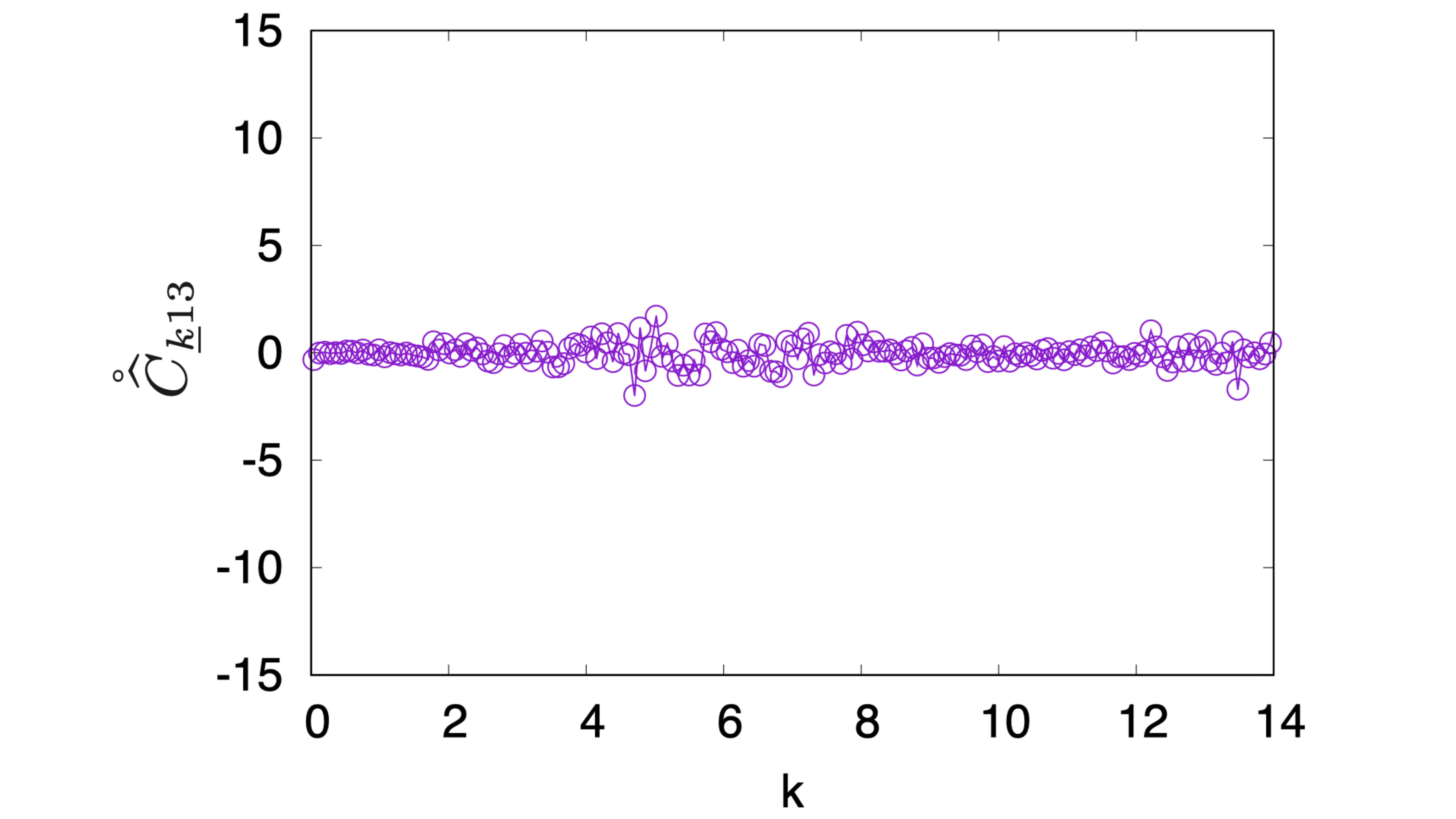}
  \caption{Angle averaged off-diagonal correlations $\tenq[2]{\mathring{\widehat{C}}}_{13 k}$. This figure
	demonstrates that the structures seen in the off-diagonal fields in Fig.~\ref{cr} are due to random numerical inaccuracies. }
	\label{offdia}
	\end{figure}
In order to show that all the fields that are expected to vanish
by symmetry are indeed zero up to numerical errors, we plot in Fig.~\ref{offdia} the angle averaged correlations of the off-diagonal fields.
Indeed, angle averaging strongly reduces the fluctuations, showing their random character.
Consequently we can safely conclude that the whole stress autocorrelation matrix
is determined solely by the pressure and torque density autocorrelation functions which are spatially isotropic.

\subsection{Long distance decay of the stress, pressure and torque autocorrelation functions}

\begin{figure}
\includegraphics[scale=0.13]{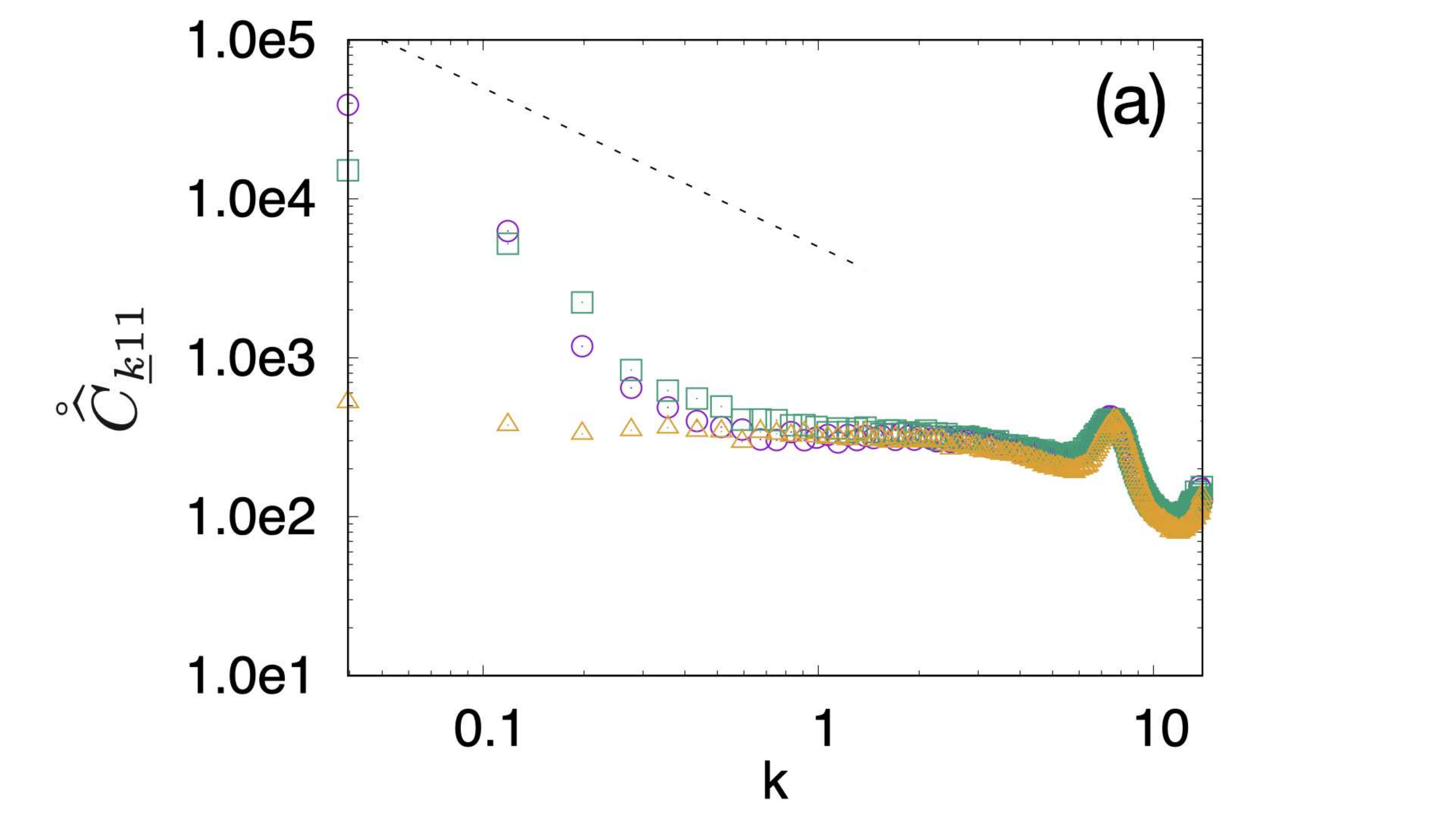}
\includegraphics[scale=0.12]{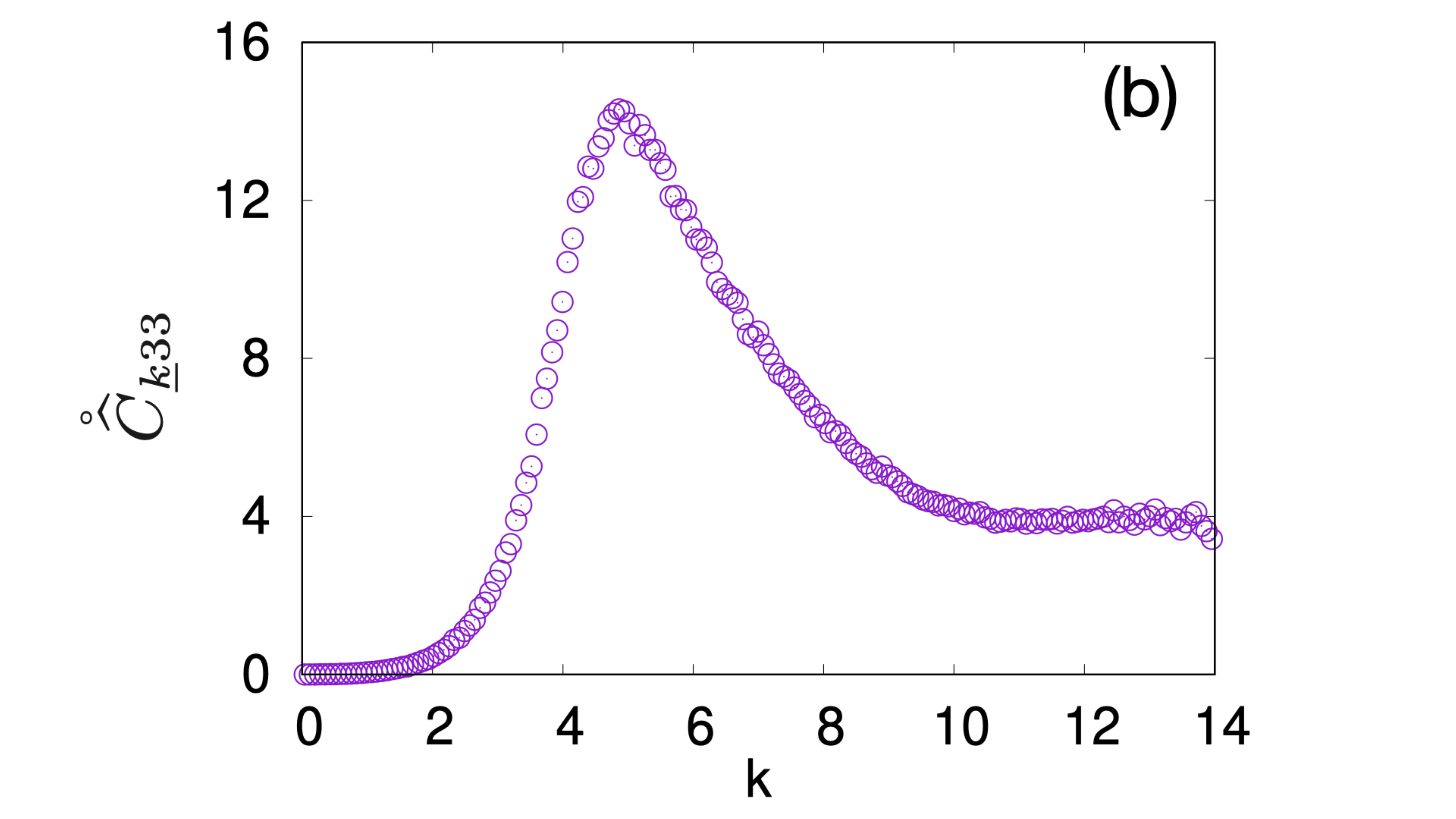}
\caption{Panel a: plot of the pressure autocorrelation function ${\mathring{\widehat{C}}}_{\vec k 11}$ vs. k. Here we show
results for $\mu=1$ for both model A (squares) and model B (circles), and for $\mu=0$ (triangles). The dashed line represents the power law $k^{-1}$.
Panel b: plot of the torque density autocorrelation function ${\mathring{\widehat{C}}}_{\vec k 33}$ vs. k.  Results are shown for Model A but model B provides essentially identical results.}
\label{pretor}
\end{figure}

The pressure and the torque angle-averaged autocorrelation functions are shown in Fig.~\ref{pretor}.
Regarding the long-distance decay, the results of our numerical simulations are quite interesting, indicating that our frictional granular matter exhibits unusual properties, very different from the friction-less counterpart. An unexpected interesting result is observed for the pressure autocorrelation function, showing a divergence at small k. In Fig.~\ref{pretor} panel a we show the results for three different models.
One is the frictional model A, where the attainment of the Coulomb law is abrupt, and another is model B where we smooth
out the approach to the Coulomb law (cf. the discussion after Eq.~\ref{Coulomb}). In both cases the systems size $N=16000$ and
$\mu=1$. Both models exhibit a similar strong divergence at $k\to 0$. For comparison, we show in the same figure the corresponding results for $\mu=0$, the friction-less case. As expected,
the friction-less case exhibits normal correlations that approach a constant value as $k\to 0$.
We estimate the exponent associated with the divergence exhibited by model A by averaging over all the components $ab=11, 12, 21, 22$ . The
result is that the data indicates a power-law divergence like $k^{-\nu}$ with $\nu$ about unity.
On the other hand, the components $ab=13, 14, 23,24,31,32,41,42$ are zero up to some randomness as seen in Fig.~\ref{offdia}.
Model B is in agreement with model A.
\subsection{The source of divergence}

\begin{figure}
 \includegraphics[scale=0.22]{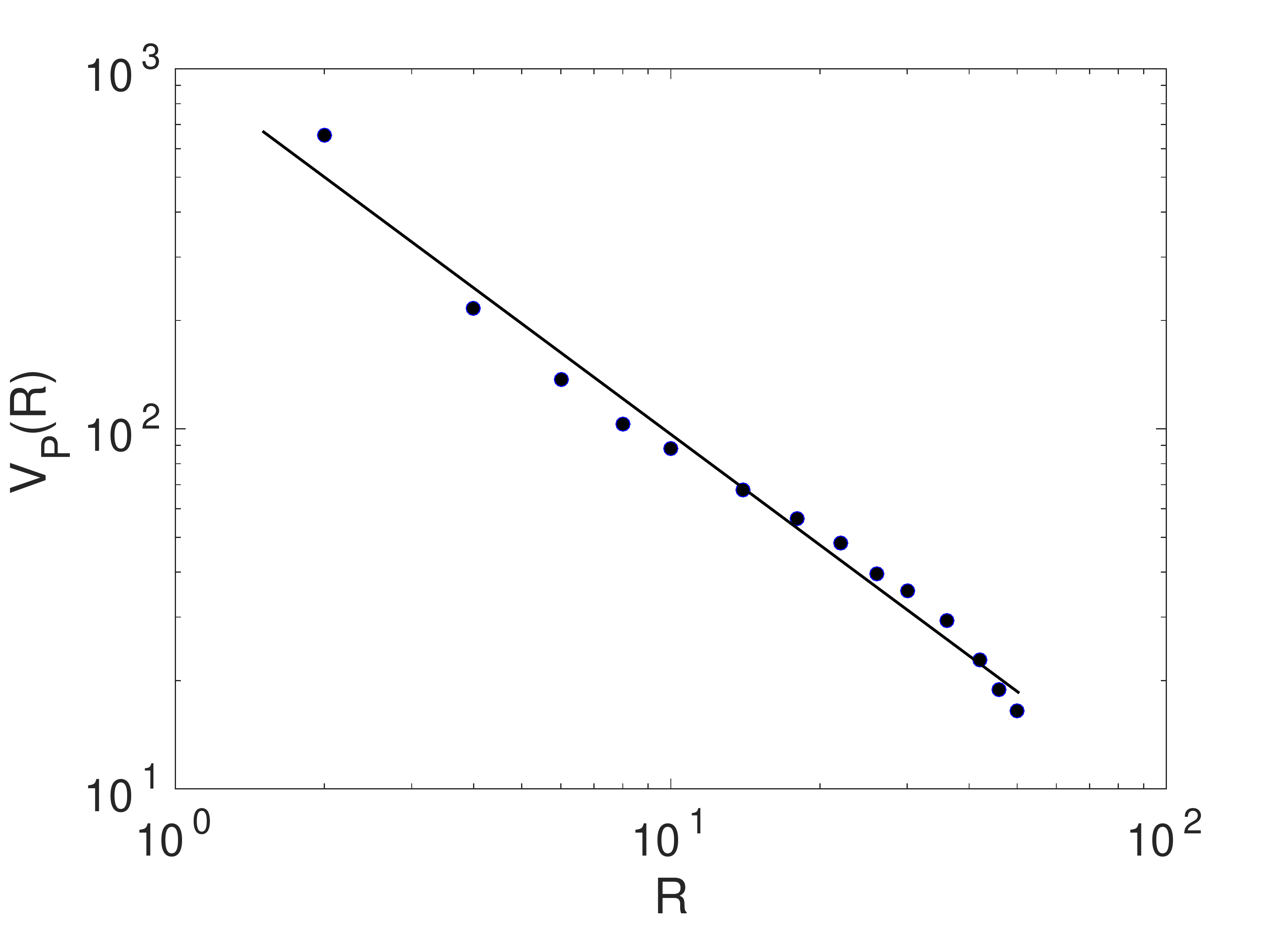}
\caption{The variance of pressures computed on circles of radius $R$ as a function of $R$. The data are shown
	as circles, the line is the best linear
fit which agrees with Eq.~(\ref{eqvar}) with $\eta$ about unity. }
\label{variance}
\end{figure}

To understand the nature of the divergence we recall that the theory guarantees that if the pressure fluctuations are normal
and the torque hyperuniform, then the asymptotics of the stress or pressure correlation as $k\to 0$ should be finite.
Accordingly we can ask which of the two, pressure or torque, is responsible for the divergences.
A very interesting and important result is in Fig.~\ref{pretor} panel b which shows the torque autocorrelation function.
The zero limit of this function as $k\to 0$ shows that the torque fluctuations are hyperuniform. In fact this
is quite intuitive: contrary to pressure, the torque has to vanish on {\bf every} disk, forcing the autocorrelation
to decay faster than normal. Since the torque fluctuations are hyperuniform, the decay of the stress autocorrelation function at large
distances is determined by the pressure statistics. We therefore measure the pressure $P(R)$ averaged on circles of radius
$R$ and compute the variance $V_P(R)$ due to circle-to-circle and sample-to-sample fluctuations:
\begin{equation}
V_P(R) \equiv \langle P(R)^2 \rangle - \langle P(R)\rangle ^2 \sim \frac{1}{R^\eta}\ ,
\label{eqvar}
\end{equation}
When the pressure has normal fluctuations this variance is expected to decay like $1/R^2$. In fact we find, cf.  Fig.~\ref{variance},  that  $V_P(R)$ decays slower,
as the power law Eq.~(\ref{eqvar}) with $\eta$ about unity. A simple calculation indicates that
\begin{equation}
\nu = 2-\eta \ ,
\end{equation}
which appears consistent.

\begin{figure}
	 \includegraphics[scale=0.25]{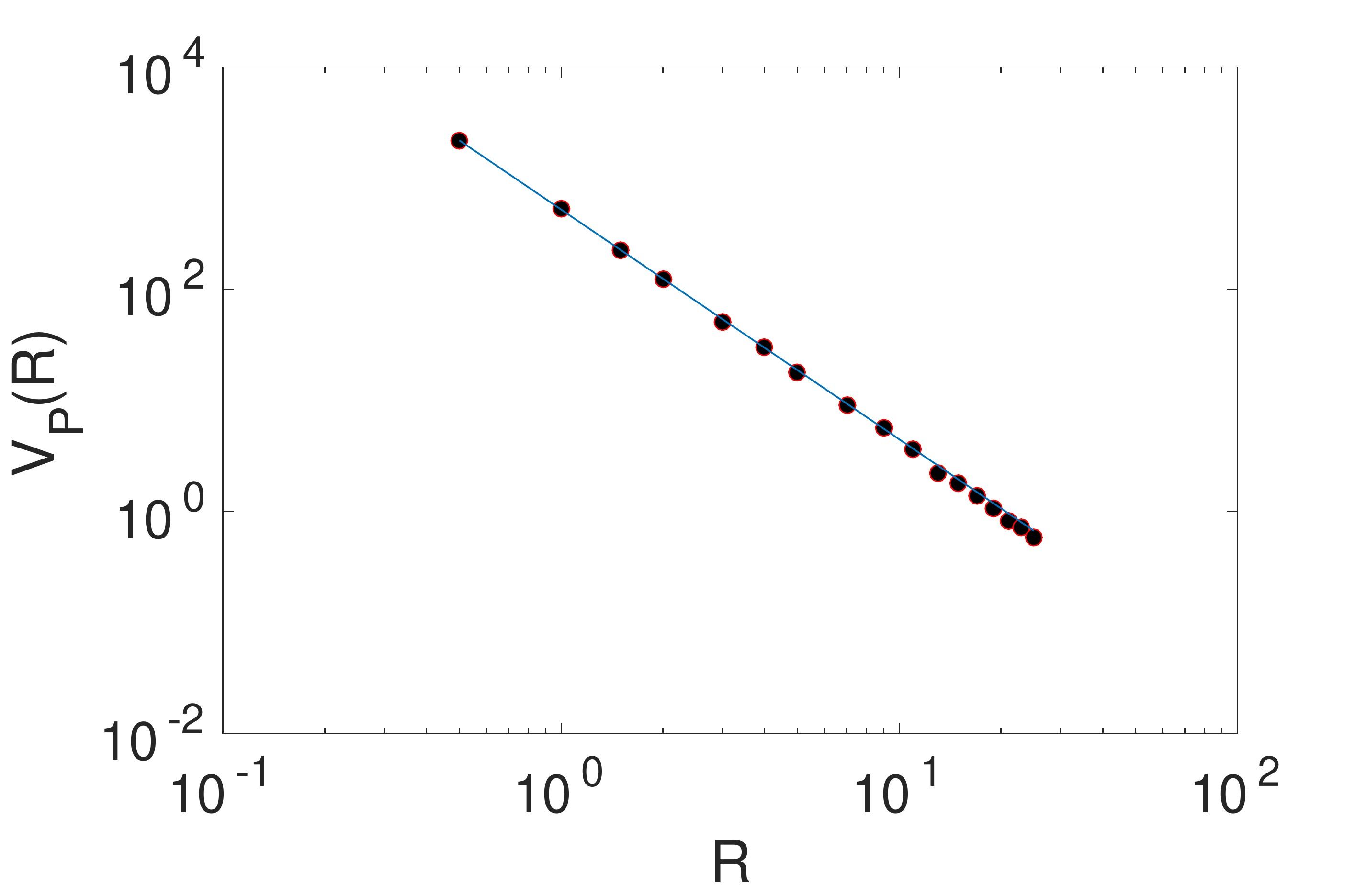}
	\caption{The variance of pressures computed on circles of radius $R$ as a function of $R$ for the system
		without friction. The linear
		fit agrees with Eq.~(\ref{eqvar}) with $\eta\approx 2$. }
	\label{varianceno}
\end{figure}

To increase our confidence in the anomalies discovered in the frictional ensembles, we repeated the very
same protocols with the very same disks but using the friction coefficient $\mu=0$. In this case we find that
the pressure fluctuations are normal, and accordingly, as the theorem proved above states, the stress
autocorrelation functions decay at large distance as expected, i.e. like $1/r^2$. The $R$ dependence of the
variance of pressure computed on circles of radius $R$ is shown in Fig.~\ref{varianceno}.
The corresponding pressure autocorrelation function as a function of $k$ is presented in the upper panel of Fig.~\ref{pretor}

We should note that the results shown in this section are at variance with the claims of Refs.\cite{17WKMT,18DeG}.
The first reference reported divergences in the $k\to 0$ limit of the pressure autocorrelation function in frictionless
samples,
and these were theoretically ``explained" in Ref. \cite{18DeG}. Our results show that the divergence
in the frictionless case is as spurious as the corresponding theoretical explanation.
\subsection{Explanation of the anomalies}
\label{explain}

At this point it is interesting to seek the physical reason for the anomalies in the pressure variance and the
consequent divergences in the autocorrelation functions. To this aim we explored the force chains in the samples
produced with and without friction. To present the force chains we compute the average magnitude of the forces
$f_{ij}$, which is denoted as $\langle f_{ij} \rangle$, and then plot all the forces whose magnitude exceeds this
average (i.e $f_{ij} \ge \langle f_{ij} \rangle$). Two typical real space maps of these force chains are shown
for two configurations compressed with the very same protocol, in panel a with friction, $\mu=1$, and in panel b
without friction. The difference is glaring: in the friction-less sample the force chains are homogeneous and
isotropic, but in the frictional sample there are clear inhomogeneities which translate to anomalous correlation
functions as observed.
\begin{figure}
   \includegraphics[scale=0.55]{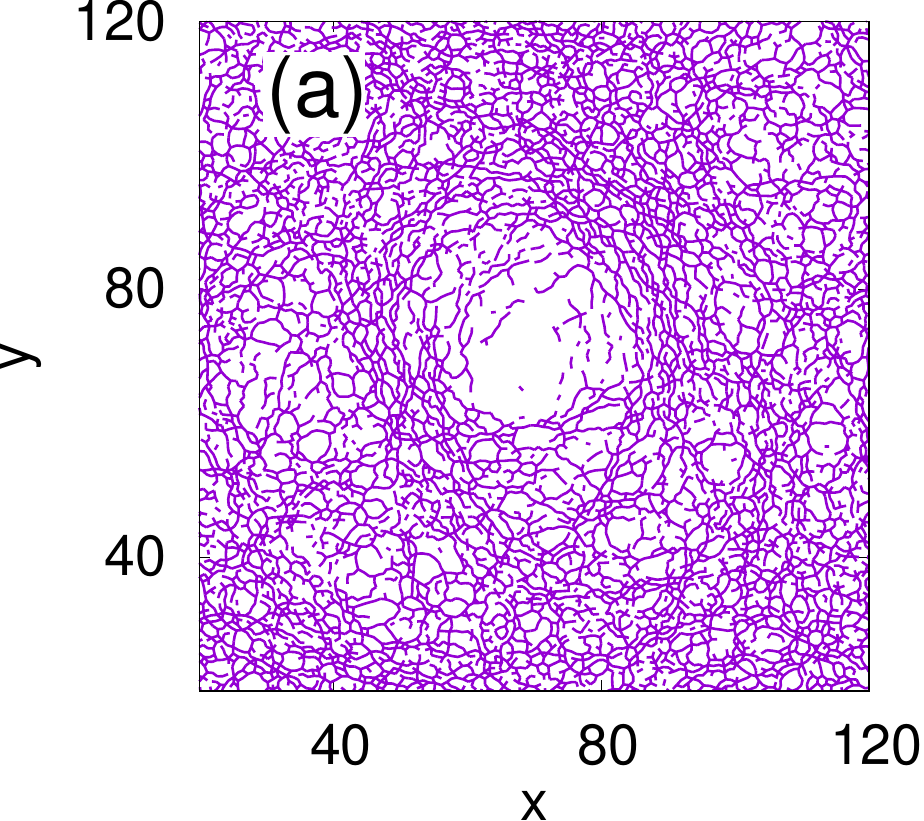}
   \includegraphics[scale=0.55]{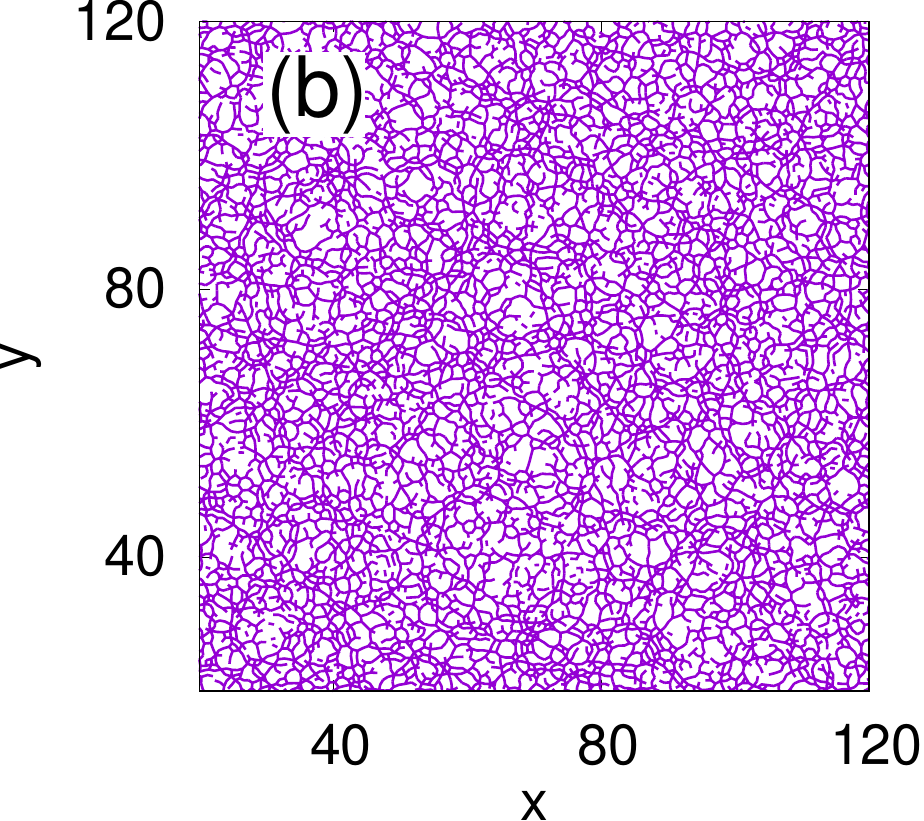}
	\caption{Force chains in compressed samples. Panel a: the frictional case, $N=16000$, $\mu=1$. Panel b:
frictionless case, $N=16000$, $\mu=0$.}
	\label{forcechains}
\end{figure}

\begin{figure}
	 \includegraphics[scale=0.55]{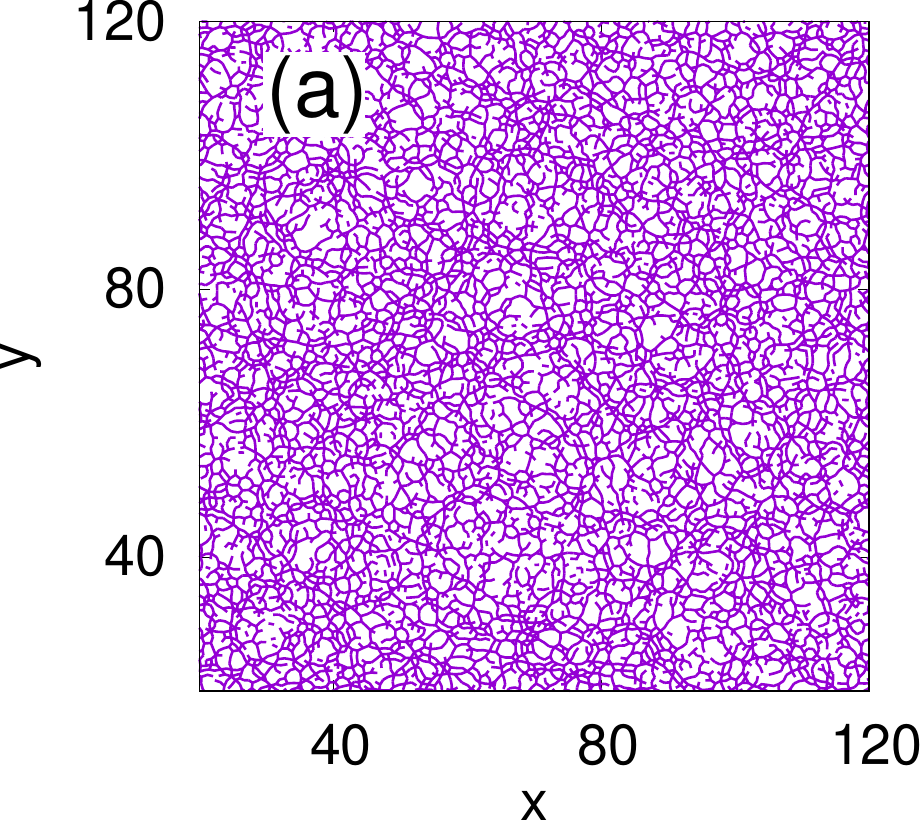}
	 \includegraphics[scale=0.12]{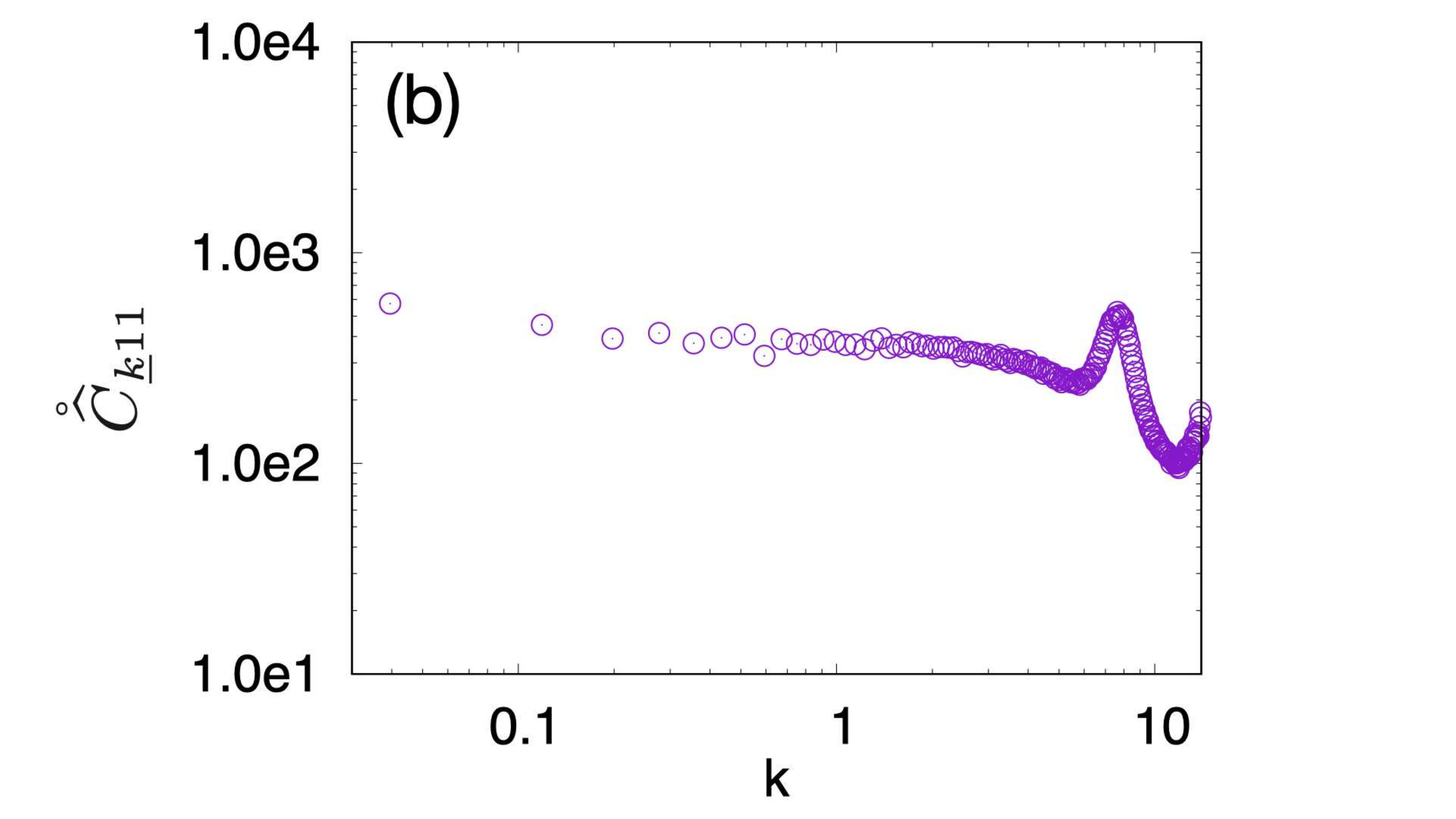}
	\caption{Panel a: force chains in a frictional compressed sample prepared by switching off the friction in the
	dilute stage before jamming. Friction was switched back on for the final compression to the target pressure. Panel
b: The correlation function  ${\mathring{\widehat{C}}}_{\vec k 11}$ vs. k. in this same sample.}
	\label{switch}
\end{figure}

A natural question then arises: when the anomalous structures of force chains get generated? Is it in the
compression stage before jamming, or in the further compression after jamming? To answer this question we switched
off the friction (i.e set $\mu=0$) in the first compression protocol before jamming, and switched back the friction
to $\mu=1$ from the point of jamming to the final attainment of the target pressure. Interestingly enough, the
anomalies disappeared. The resulting force chains and autocorrelation function are shown in Fig.~\ref{switch}
in panels a and b respectively.
We can therefore conclude that the anomalous correlations in force chains are created already in the
dilute regime before jamming. Once the frictional system jams, these correlations cannot be removed during
the additional compression to the target pressure!

It is interesting to note that the inverse experiment, in which the compression until jamming is done with
friction on, switching off the friction for the further compression to the target pressure, is not a useful
exercise. The reason is that jamming in frictional matter occurs at a lower area fraction than
in friction-less samples. Therefore once friction is put to zero upon jamming, the system gets unjammed,
and there is a stage of further compression until friction-less jamming. In this stage all the anomalous
correlations built during the frictional compression disappear, leading to normal pressure fluctuations
in the compressed sample.

\section{Summary and Conclusions}
\label{summary}

Frictional granular matter is all around us, and the tendency over the years was to assume that granular samples exhibit
``normal" elastic behavior typical to amorphous solids. In this paper we focused on the long-distance decay of autocorrelation
functions of various components of the stress tensor. The presence of friction was shown to distinguish these materials
from amorphous solids in which the microscopic interaction are Hamiltonian and the constituents do not experience torques.
In the frictionless case granular systems are expected to exhibit normal decay at long distances, with a tail that is
typical to the elastic Green's function. In fact, it was proven that it is sufficient that the pressure exhibits normal
fluctuations to guarantee normal decay.  Once friction is added, normal pressure fluctuations are not sufficient,
in addition one needs to guarantee that the torque fluctuations are hyper uniform. We have shown in this paper that with these two  conditions satisfied, the long distance decay of stress correlation function conforms with elastic Green's functions. To test the predictions
of the theory we set up numerical simulations of frictional granular matter using the standard Kundall-Strack model of
normal and tangential forces. The tangential forces are limited as usual by the Coulomb law Eq.~(\ref{Coulomb}).  We examined
two models of the approach to the Coulomb limit, one abrupt (model A)  and one smooth, with two derivatives, model B.
All our simulation results agreed for the two models.

The most striking result of our simulations was that the compressed frictional granular matter exhibited divergences
in the $k\to 0$ limit of the relevant stress autocorrelation functions. Normal behavior like the elastic Green's function
is consistent with these functions going to a constant in this limit. In light of the theory presented above, the failure
to conform with elasticity theory must be related to either the pressure or the torque having unusual properties.
The numerical simulations confirmed that the torque is hyperuniform as expected. The culprit had to be the pressure,
and indeed testing the fluctuations of the pressure we discovered that its variance decays anomalously slowly
with the area, busting one of the conditions for normal decay. Having discovered this, it became important to find
when and how anomalous pressure correlations were produced in the materials. The protocol of compression started with
a dilute system of zero pressure, that was compressed isotropically until the system jammed, and then further compression brought it
to a target pressure. We have discovered that the anomalous correlations form in the dilute phase, while the pressure
was still zero. Once the system jammed these anomalous correlations were already imprinted in the material and could not
be released. The signature is seen in the force chains that remain inhomogeneous while compressing after jamming.

Switching friction off in the dilute phase removes the anomalies, even when we switch the friction back on after
jamming, in the final compression. Of course, this is a numerical trick that cannot be done in a physical system
of frictional granules. The crucial question that this study underlines is therefore {\bf ``is this anomalous behavior
generic to frictional granular matter or is it a consequence of the class of models employed".} The answer to
this exciting question must await similar physical experiments and/or simulations with fundamentally different
models of frictional granular matter. Both of these are tasks for the future.

\section{Acknowledgements}
This work had been supported in part by the US-Israel Binational Science Foundation and by the scientific and cooperation agreement between Italy and Israel through the project COMPAMP/DISORDER.

\bibliography{All}

\begin{thebibliography}{19}%
\makeatletter
\providecommand \@ifxundefined [1]{%
 \@ifx{#1\undefined}
}%
\providecommand \@ifnum [1]{%
 \ifnum #1\expandafter \@firstoftwo
 \else \expandafter \@secondoftwo
 \fi
}%
\providecommand \@ifx [1]{%
 \ifx #1\expandafter \@firstoftwo
 \else \expandafter \@secondoftwo
 \fi
}%
\providecommand \natexlab [1]{#1}%
\providecommand \enquote  [1]{``#1''}%
\providecommand \bibnamefont  [1]{#1}%
\providecommand \bibfnamefont [1]{#1}%
\providecommand \citenamefont [1]{#1}%
\providecommand \href@noop [0]{\@secondoftwo}%
\providecommand \href [0]{\begingroup \@sanitize@url \@href}%
\providecommand \@href[1]{\@@startlink{#1}\@@href}%
\providecommand \@@href[1]{\endgroup#1\@@endlink}%
\providecommand \@sanitize@url [0]{\catcode `\\12\catcode `\$12\catcode
  `\&12\catcode `\#12\catcode `\^12\catcode `\_12\catcode `\%12\relax}%
\providecommand \@@startlink[1]{}%
\providecommand \@@endlink[0]{}%
\providecommand \url  [0]{\begingroup\@sanitize@url \@url }%
\providecommand \@url [1]{\endgroup\@href {#1}{\urlprefix }}%
\providecommand \urlprefix  [0]{URL }%
\providecommand \Eprint [0]{\href }%
\providecommand \doibase [0]{http://dx.doi.org/}%
\providecommand \selectlanguage [0]{\@gobble}%
\providecommand \bibinfo  [0]{\@secondoftwo}%
\providecommand \bibfield  [0]{\@secondoftwo}%
\providecommand \translation [1]{[#1]}%
\providecommand \BibitemOpen [0]{}%
\providecommand \bibitemStop [0]{}%
\providecommand \bibitemNoStop [0]{.\EOS\space}%
\providecommand \EOS [0]{\spacefactor3000\relax}%
\providecommand \BibitemShut  [1]{\csname bibitem#1\endcsname}%
\let\auto@bib@innerbib\@empty
\bibitem [{\citenamefont {Henkes}\ and\ \citenamefont
  {Chakraborty}(2009)}]{09HC}%
  \BibitemOpen
  \bibfield  {author} {\bibinfo {author} {\bibfnamefont {S.}~\bibnamefont
  {Henkes}}\ and\ \bibinfo {author} {\bibfnamefont {B.}~\bibnamefont
  {Chakraborty}},\ }\href {\doibase 10.1103/PhysRevE.79.061301} {\bibfield
  {journal} {\bibinfo  {journal} {Phys. Rev. E}\ }\textbf {\bibinfo {volume}
  {79}},\ \bibinfo {pages} {061301} (\bibinfo {year} {2009})}\BibitemShut
  {NoStop}%
\bibitem [{\citenamefont {Lema\^{\i}tre}(2014)}]{14Lem}%
  \BibitemOpen
  \bibfield  {author} {\bibinfo {author} {\bibfnamefont {A.}~\bibnamefont
  {Lema\^{\i}tre}},\ }\href {\doibase 10.1103/PhysRevLett.113.245702}
  {\bibfield  {journal} {\bibinfo  {journal} {Phys. Rev. Lett.}\ }\textbf
  {\bibinfo {volume} {113}},\ \bibinfo {pages} {245702} (\bibinfo {year}
  {2014})}\BibitemShut {NoStop}%
\bibitem [{\citenamefont {Lema\^{\i}tre}(2015)}]{15Lem}%
  \BibitemOpen
  \bibfield  {author} {\bibinfo {author} {\bibfnamefont {A.}~\bibnamefont
  {Lema\^{\i}tre}},\ }\href {\doibase 10.1063/1.4933235} {\bibfield  {journal}
  {\bibinfo  {journal} {The Journal of Chemical Physics}\ }\textbf {\bibinfo
  {volume} {143}},\ \bibinfo {pages} {164515} (\bibinfo {year}
  {2015})}\BibitemShut {NoStop}%
\bibitem [{\citenamefont {Lema\^{\i}tre}(2017)}]{17Lem}%
  \BibitemOpen
  \bibfield  {author} {\bibinfo {author} {\bibfnamefont {A.}~\bibnamefont
  {Lema\^{\i}tre}},\ }\href {\doibase 10.1103/PhysRevE.96.052101} {\bibfield
  {journal} {\bibinfo  {journal} {Phys. Rev. E}\ }\textbf {\bibinfo {volume}
  {96}},\ \bibinfo {pages} {052101} (\bibinfo {year} {2017})}\BibitemShut
  {NoStop}%
\bibitem [{\citenamefont {Lema\^{\i}tre}(2018)}]{18Lem}%
  \BibitemOpen
  \bibfield  {author} {\bibinfo {author} {\bibfnamefont {A.}~\bibnamefont
  {Lema\^{\i}tre}},\ }\href {\doibase 10.1063/1.5041461} {\bibfield  {journal}
  {\bibinfo  {journal} {The Journal of Chemical Physics}\ }\textbf {\bibinfo
  {volume} {149}},\ \bibinfo {pages} {104107} (\bibinfo {year} {2018})},\
  \Eprint {http://arxiv.org/abs/https://doi.org/10.1063/1.5041461}
  {https://doi.org/10.1063/1.5041461} \BibitemShut {NoStop}%
\bibitem [{\citenamefont {Edwards}\ and\ \citenamefont
  {Oakeshott}(1989)}]{89EO}%
  \BibitemOpen
  \bibfield  {author} {\bibinfo {author} {\bibfnamefont {S.}~\bibnamefont
  {Edwards}}\ and\ \bibinfo {author} {\bibfnamefont {R.}~\bibnamefont
  {Oakeshott}},\ }\href {\doibase https://doi.org/10.1016/0378-4371(89)90034-4}
  {\bibfield  {journal} {\bibinfo  {journal} {Physica A: Statistical Mechanics
  and its Applications}\ }\textbf {\bibinfo {volume} {157}},\ \bibinfo {pages}
  {1080 } (\bibinfo {year} {1989})}\BibitemShut {NoStop}%
\bibitem [{\citenamefont {Cundall}\ and\ \citenamefont {Strack}(1979)}]{79CS}%
  \BibitemOpen
  \bibfield  {author} {\bibinfo {author} {\bibfnamefont {P.~A.}\ \bibnamefont
  {Cundall}}\ and\ \bibinfo {author} {\bibfnamefont {O.~D.~L.}\ \bibnamefont
  {Strack}},\ }\href {\doibase 10.1680/geot.1979.29.1.47} {\bibfield  {journal}
  {\bibinfo  {journal} {Géotechnique}\ }\textbf {\bibinfo {volume} {29}},\
  \bibinfo {pages} {47} (\bibinfo {year} {1979})}\BibitemShut {NoStop}%
\bibitem [{\citenamefont {Silbert}\ \emph {et~al.}(2001)\citenamefont
  {Silbert}, \citenamefont {Erta\ifmmode~\mbox{\c{s}}\else \c{s}\fi{}},
  \citenamefont {Grest}, \citenamefont {Halsey}, \citenamefont {Levine},\ and\
  \citenamefont {Plimpton}}]{01SEGHLP}%
  \BibitemOpen
  \bibfield  {author} {\bibinfo {author} {\bibfnamefont {L.~E.}\ \bibnamefont
  {Silbert}}, \bibinfo {author} {\bibfnamefont {D.}~\bibnamefont
  {Erta\ifmmode~\mbox{\c{s}}\else \c{s}\fi{}}}, \bibinfo {author}
  {\bibfnamefont {G.~S.}\ \bibnamefont {Grest}}, \bibinfo {author}
  {\bibfnamefont {T.~C.}\ \bibnamefont {Halsey}}, \bibinfo {author}
  {\bibfnamefont {D.}~\bibnamefont {Levine}}, \ and\ \bibinfo {author}
  {\bibfnamefont {S.~J.}\ \bibnamefont {Plimpton}},\ }\href {\doibase
  10.1103/PhysRevE.64.051302} {\bibfield  {journal} {\bibinfo  {journal} {Phys.
  Rev. E}\ }\textbf {\bibinfo {volume} {64}},\ \bibinfo {pages} {051302}
  (\bibinfo {year} {2001})}\BibitemShut {NoStop}%
\bibitem [{\citenamefont {Chattoraj}\ \emph
  {et~al.}(2019{\natexlab{a}})\citenamefont {Chattoraj}, \citenamefont
  {Gendelman}, \citenamefont {Pica~Ciamarra},\ and\ \citenamefont
  {Procaccia}}]{19CGPP}%
  \BibitemOpen
  \bibfield  {author} {\bibinfo {author} {\bibfnamefont {J.}~\bibnamefont
  {Chattoraj}}, \bibinfo {author} {\bibfnamefont {O.}~\bibnamefont
  {Gendelman}}, \bibinfo {author} {\bibfnamefont {M.}~\bibnamefont
  {Pica~Ciamarra}}, \ and\ \bibinfo {author} {\bibfnamefont {I.}~\bibnamefont
  {Procaccia}},\ }\href@noop {} {\bibfield  {journal} {\bibinfo  {journal}
  {Phys. Rev. Lett.}\ }\textbf {\bibinfo {volume} {123}},\ \bibinfo {pages}
  {098003} (\bibinfo {year} {2019}{\natexlab{a}})}\BibitemShut {NoStop}%
\bibitem [{\citenamefont {Chattoraj}\ \emph
  {et~al.}(2019{\natexlab{b}})\citenamefont {Chattoraj}, \citenamefont
  {Gendelman}, \citenamefont {Ciamarra},\ and\ \citenamefont
  {Procaccia}}]{19CGPPa}%
  \BibitemOpen
  \bibfield  {author} {\bibinfo {author} {\bibfnamefont {J.}~\bibnamefont
  {Chattoraj}}, \bibinfo {author} {\bibfnamefont {O.}~\bibnamefont
  {Gendelman}}, \bibinfo {author} {\bibfnamefont {M.~P.}\ \bibnamefont
  {Ciamarra}}, \ and\ \bibinfo {author} {\bibfnamefont {I.}~\bibnamefont
  {Procaccia}},\ }\href {\doibase 10.1103/PhysRevE.100.042901} {\bibfield
  {journal} {\bibinfo  {journal} {Phys. Rev. E}\ }\textbf {\bibinfo {volume}
  {100}},\ \bibinfo {pages} {042901} (\bibinfo {year}
  {2019}{\natexlab{b}})}\BibitemShut {NoStop}%
\bibitem [{\citenamefont {Bonfanti}\ \emph {et~al.}(2020)\citenamefont
  {Bonfanti}, \citenamefont {Chattoraj}, \citenamefont {Guerra}, \citenamefont
  {Procaccia},\ and\ \citenamefont {Zapperi}}]{20BCGPZ}%
  \BibitemOpen
  \bibfield  {author} {\bibinfo {author} {\bibfnamefont {S.}~\bibnamefont
  {Bonfanti}}, \bibinfo {author} {\bibfnamefont {J.}~\bibnamefont {Chattoraj}},
  \bibinfo {author} {\bibfnamefont {R.}~\bibnamefont {Guerra}}, \bibinfo
  {author} {\bibfnamefont {I.}~\bibnamefont {Procaccia}}, \ and\ \bibinfo
  {author} {\bibfnamefont {S.}~\bibnamefont {Zapperi}},\ }\href {\doibase
  10.1103/PhysRevE.101.052902} {\bibfield  {journal} {\bibinfo  {journal}
  {Phys. Rev. E}\ }\textbf {\bibinfo {volume} {101}},\ \bibinfo {pages}
  {052902} (\bibinfo {year} {2020})}\BibitemShut {NoStop}%
\bibitem [{\citenamefont {Plimpton}(1995)}]{95Pli}%
  \BibitemOpen
  \bibfield  {author} {\bibinfo {author} {\bibfnamefont {S.}~\bibnamefont
  {Plimpton}},\ }\href {\doibase 10.1006/jcph.1995.1039} {\bibfield  {journal}
  {\bibinfo  {journal} {Journal of Computational Physics}\ }\textbf {\bibinfo
  {volume} {117}},\ \bibinfo {pages} {1} (\bibinfo {year} {1995})}\BibitemShut
  {NoStop}%
\bibitem [{\citenamefont {Kloss}(2012)}]{12KGHAP}%
  \BibitemOpen
  \bibfield  {author} {\bibinfo {author} {\bibfnamefont {C.}~\bibnamefont
  {Kloss}},\ }\href@noop {} {\bibfield  {journal} {\bibinfo  {journal} {Prog.
  Comput. Fluid Dyn.}\ }\textbf {\bibinfo {volume} {12}},\ \bibinfo {pages}
  {140} (\bibinfo {year} {2012})}\BibitemShut {NoStop}%
\bibitem [{\citenamefont {Goldhirsch}\ and\ \citenamefont
  {Goldenberg}(2002)}]{02GG}%
  \BibitemOpen
  \bibfield  {author} {\bibinfo {author} {\bibfnamefont {I.}~\bibnamefont
  {Goldhirsch}}\ and\ \bibinfo {author} {\bibfnamefont {C.}~\bibnamefont
  {Goldenberg}},\ }\href@noop {} {\bibfield  {journal} {\bibinfo  {journal}
  {The European Physical Journal E}\ }\textbf {\bibinfo {volume} {9}},\
  \bibinfo {pages} {245} (\bibinfo {year} {2002})}\BibitemShut {NoStop}%
\bibitem [{\citenamefont {Evans}\ \emph {et~al.}(1990)\citenamefont {Evans},
  \citenamefont {Cohen},\ and\ \citenamefont {Morriss}}]{90ECM}%
  \BibitemOpen
  \bibfield  {author} {\bibinfo {author} {\bibfnamefont {D.~J.}\ \bibnamefont
  {Evans}}, \bibinfo {author} {\bibfnamefont {E.~G.~D.}\ \bibnamefont {Cohen}},
  \ and\ \bibinfo {author} {\bibfnamefont {G.~P.}\ \bibnamefont {Morriss}},\
  }\href {\doibase 10.1103/PhysRevA.42.5990} {\bibfield  {journal} {\bibinfo
  {journal} {Phys. Rev. A}\ }\textbf {\bibinfo {volume} {42}},\ \bibinfo
  {pages} {5990} (\bibinfo {year} {1990})}\BibitemShut {NoStop}%
\bibitem [{\citenamefont {Riesz}(1949)}]{Riesz1949}%
  \BibitemOpen
  \bibfield  {author} {\bibinfo {author} {\bibfnamefont {M.}~\bibnamefont
  {Riesz}},\ }\href {\doibase 10.1007/BF02395016} {\bibfield  {journal}
  {\bibinfo  {journal} {Acta Mathematica}\ }\textbf {\bibinfo {volume} {81}},\
  \bibinfo {pages} {1} (\bibinfo {year} {1949})}\BibitemShut {NoStop}%
\bibitem [{\citenamefont {Landkof}(1972)}]{Landkof1972}%
  \BibitemOpen
  \bibfield  {author} {\bibinfo {author} {\bibfnamefont {N.}~\bibnamefont
  {Landkof}},\ }\href@noop {} {\emph {\bibinfo {title} {Foundations of modern
  potential theory}}},\ \bibinfo {number} {MR0350027}\ (\bibinfo  {publisher}
  {Springer-Verlag},\ \bibinfo {address} {Berlin, New York},\ \bibinfo {year}
  {1972})\BibitemShut {NoStop}%
\bibitem [{\citenamefont {Wu}\ \emph {et~al.}(2017)\citenamefont {Wu},
  \citenamefont {Karimi}, \citenamefont {Maloney},\ and\ \citenamefont
  {Teitel}}]{17WKMT}%
  \BibitemOpen
  \bibfield  {author} {\bibinfo {author} {\bibfnamefont {Y.}~\bibnamefont
  {Wu}}, \bibinfo {author} {\bibfnamefont {K.}~\bibnamefont {Karimi}}, \bibinfo
  {author} {\bibfnamefont {C.~E.}\ \bibnamefont {Maloney}}, \ and\ \bibinfo
  {author} {\bibfnamefont {S.}~\bibnamefont {Teitel}},\ }\href {\doibase
  10.1103/PhysRevE.96.032902} {\bibfield  {journal} {\bibinfo  {journal} {Phys.
  Rev. E}\ }\textbf {\bibinfo {volume} {96}},\ \bibinfo {pages} {032902}
  (\bibinfo {year} {2017})}\BibitemShut {NoStop}%
\bibitem [{\citenamefont {DeGiuli}(2018)}]{18DeG}%
  \BibitemOpen
  \bibfield  {author} {\bibinfo {author} {\bibfnamefont {E.}~\bibnamefont
  {DeGiuli}},\ }\href {\doibase 10.1103/PhysRevE.98.033001} {\bibfield
  {journal} {\bibinfo  {journal} {Phys. Rev. E}\ }\textbf {\bibinfo {volume}
  {98}},\ \bibinfo {pages} {033001} (\bibinfo {year} {2018})}\BibitemShut
  {NoStop}%
\end{thebibliography}%

\end{document}